\documentclass{article}

\let\origS\S
\usepackage[utf8]{inputenc}
\usepackage{ae,aecompl,array, amsmath, amssymb, amsthm, color, complexity, verbatim, bbm,amsfonts, hyperref, mathrsfs, graphics,tikz}\usepackage{stmaryrd}
\usepackage[margin=1.333in]{geometry}

\let\S\origS

\newcommand{\CC}{\ensuremath{\mathscr{C}}}
\newcommand{\CS}{\ensuremath{\mathscr{S}}}

\renewcommand{\C}{\mathcal{C}}
\newcommand{\Dc}{\mathcal{D}}
\newcommand{\Ec}{\mathcal{E}}

\newcommand{\Xc}{\mathcal{S}}

\newcommand{\hh}{\widehat{h}}

\newcommand{\SRSC}{\ensuremath{\mathrm{S}\textrm{-}\mathrm{RSC }}}

\newcommand{\CSP}{\ensuremath{\mathrm{CS}}}

\newcommand{\Comp}{\mathbb{C}}

\newcommand{\F}{\mathbb{F}}
\newcommand{\Z}{\mathbb{Z}}
\newcommand{\Fq}{\F_q}
\newcommand{\Fqn}{\F_q^n}
\newcommand{\Zq}{\Z_q}

\newcommand{\fq}{\Fq}

\newcommand{\Iint}[2]{\llbracket #1 , #2 \rrbracket}

\newcommand{\fsRS}[1]{\text{\bf RS}_{#1}}
\newcommand{\RS}[2]{\text{\bf RS}_{#1}(#2)}
\newcommand{\GRS}[3]{\text{\bf GRS}_{#1}(#2,#3)}

\newcommand{\SD}{\mathrm{SD}}

\newcommand{\GOOD}{\mathrm{Good}}
\newcommand{\negl}{\mathrm{negl}}
\newcommand{\Ideal}{\mathrm{ideal}}
\newcommand{\Real}{\mathrm{real}}
\renewcommand{\aa}{\mathcal{A}}

\newcommand{\hF}{\widehat{F}}

\DeclareMathOperator{\tr}{tr}

\newcommand{\trsp}[1]{{#1}^{\intercal}}

\newcommand{\eqdef}{\mathop{=}\limits^{\triangle}}
\newcommand{\Unif}{\leftarrow}

\newcommand{\ie}{\textit{i.e.}}
\newcommand{\eps}{\varepsilon}

\newcommand{\COMMENT}[1]{}

\newcommand{\SAINTERPOL}{\ensuremath{$\cS$\textrm{-Polynomial Interpolation}}}
\newcommand{\ISIS}{\mbox{ISIS}}

\newcommand{\RSISIS}{\mathrm{RS}\textrm{-}\mathrm{ISIS}}

\newcommand{\esp}{\mathbb{E}}
\DeclareMathOperator{\var}{\mathbf{Var}}

\newcommand{\prob}{\Pr}

\newcommand{\ket}[1]{|#1\rangle}

\newcommand{\braket}[2]{\langle #1 | #2 \rangle}

\newcommand{\norm}[1]{\ensuremath{\lvert\!\lvert #1 \rvert\!\rvert}}

\newcommand{\SIS}{\mbox{SIS}}

\newcommand{\one}{\mathbbm{1}}

\DeclareMathOperator{\QFTt}{QFT}
\newcommand{\carf}[1]{\chi_{#1}}
\newcommand{\car}[2]{\chi_{#1}\left(#2 \right)}
\newcommand{\hf}{\widehat{f}}
\newcommand{\hg}{\widehat{g}}

\newcommand{\QFT}[1]{\widehat{#1}}
\newcommand{\FT}[1]{\QFT{#1}}

\newcommand{\Rbra}[1]{\left( #1 \right)} 

\newcommand{\mat}[1]{\ensuremath{\boldsymbol{#1}}}

\makeatletter
\newcommand{\ostar}{\mathbin{\mathpalette\make@circled\star}}
\newcommand{\make@circled}[2]{%
	\ooalign{$\m@th#1\smallbigcirc{#1}$\cr\hidewidth$\m@th#1#2$\hidewidth\cr}%
}
\newcommand{\smallbigcirc}[1]{%
	\vcenter{\hbox{\scalebox{0.77778}{$\m@th#1\bigcirc$}}}%
}
\makeatother

\newcommand{\cv}{{\mat{c}}}

\newcommand{\ev}{\mat{e}}

\newcommand{\sv}{{\mat{s}}}
\newcommand{\uv}{\mat{u}}
\newcommand{\vv}{\mat{v}}
\newcommand{\xv}{{\mat{x}}}
\newcommand{\yv}{{\mat{y}}}

\newcommand{\zero}{\mathbf{0}}

\newcommand{\Gm}{\ensuremath{\mathbf{G}}}
\newcommand{\Hm}{\ensuremath{\mathbf{H}}}

\renewcommand{\E}{\mathbb{E}}

\newcommand{\zerov}{\mathbf{0}}

\newcommand{\OO}[1]{O\Rbra{#1}}

\DeclareMathOperator{\Supp}{Supp}

\newtheorem{theorem}{Theorem}

\newtheorem{definition}{Definition}
\newtheorem{lemma}{Lemma}
\newtheorem{proposition}{Proposition}
\newtheorem{corollary}{Corollary}

\newtheorem{problem}{Problem}
\newtheorem{fact}{Fact}

\renewcommand{\and}{\mbox{ and }}

\newcommand{\cadre}[1]
{
	\begin{tabular}{|p{0.99\textwidth}|}
		\hline
		\vspace*{-0.3cm}
		#1 \\ $ \ $ \\
		\hline
	\end{tabular}
	
}

\title{Quantum advantage from soft decoders}
\author{André Chailloux and Jean-Pierre Tillich \\ Inria de Paris, COSMIQ team \\ \texttt{andre.chailloux@inria.fr} \quad \texttt{jean-pierre.tillich@inria.fr}}

\begin{document}
\maketitle
\begin{abstract}
	In the last years, Regev's reduction has been used as a quantum algorithmic tool for providing a quantum advantage for variants of the decoding problem. Following this line of work, the authors of \cite{JSW+24} have recently come up with a quantum algorithm called ``decoded quantum interferometry'' that is able to solve in polynomial time  several optimization problems. They study in particular the Optimal Polynomial Interpolation (OPI) problem, which can be seen as a decoding problem on Reed-Solomon codes. 
	
	In this work, we provide strong improvements for some instantiations of the OPI problem. The most notable improvements are for the $\ISIS_{\infty}$ problem (originating from lattice-based cryptography) on Reed-Solomon codes
	 but we also study different constraints for OPI. Our results provide natural and convincing decoding problems for which we believe to have a quantum advantage.
	
	Our proof techniques involve the use of a \emph{soft decoder} for Reed-Solomon codes, namely the decoding algorithm from Koetter and Vardy~\cite{KV03}. In order to be able to use this decoder in the setting of Regev's reduction, we provide a novel generic reduction from a syndrome decoding problem to a coset sampling problem, providing a powerful and simple to use theorem, which generalizes previous work and is of independent interest. We also provide an extensive study of OPI using the Koetter and Vardy algorithm.

\end{abstract}
\section{Introduction}
\subsection{Context}

In the last years, Regev's reduction has been used as a quantum algorithmic tool for providing a quantum advantages for variants of decoding problems.  Chen, Liu and Zhandry~\cite{CLZ22} initiated this line of research and showed how to use Regev's reduction for algorithmic purposes. They solved in quantum polynomial time the $\SIS_{\infty}$ problem for some parameters. While the parameters where this quantum algorithm works in polynomial time are somewhat extreme, there is no known classical algorithm that solves $\SIS_{\infty}$ for these parameters. Then, Yamakawa and Zhandry~\cite{YZ24} proved their algorithm for quantum advantage without structure using essentially a decoding problem. 
\paragraph{Quantum advantage with ``decoded quantum interferometry''.}
Recently, \cite{JSW+24} has come up with a quantum algorithm called ``decoded quantum interferometry'' that is able to solve in polynomial time  several optimization problems.
One of them is sparse max-XORSAT and the other one is Optimal Polynomial Interpolation (OPI). The quantum algorithm achieves on certain instances of the first problem a better approximation ratio
on average than simulated annealing. However, there are specialized classical algorithms tailored to this problem which provide a better approximation ratio. Much more intriguing is the
second problem, namely OPI. It basically consists for a finite field $\Fq$ in finding a polynomial in $\Fq[X]$ of some degree less than some bound $k$ which satisfies as many intersection constraints as possible. Here
there are as many intersection constraints as there are non zero elements $x$ in $\Fq$. Each constraint corresponds to
a set $\cS_x \subset \Fq$ and the intersection constraint at $x$ just means that $P(x)$ should be in $\cS_x$. To illustrate the quantum speedup which can be achieved, the authors of \cite{JSW+24} consider the case of balanced sets (the size of the $\cS_x$'s is $\approx \frac{q}{2}$), the bound on the degree is $k = \frac{q}{10}$. In this case, their quantum algorithm is able to produce in polynomial time polynomials of degree $\leq d$ which satisfy a fraction $\frac{1}{2} + \frac{\sqrt{19}}{20} \approx 0.7179$ of constraints. Remarkably the best known polynomial time classical algorithm achieves to satisfy only a fraction
$0.55$ of the constraints. As the authors note in \cite{JSW+24}, the margin of the quantum advantage for the approximation fraction is satisfyingly large.

Both problems are actually an instance of a decoding problem which appears in rate distortion theory. We are given a linear code $\C$, \ie\ a subspace of $\F_q^n$ of some dimension $k$ and a word
$\yv=(y_i)_{1 \leq i \leq n} \in \Fqn$ and we are asked to find a codeword $\cv=(c_i)_{1 \leq i \leq n}$ (an element of $\C$) which is sufficiently close to $\yv$, where the distance $d(\cdot,\cdot)$ is say some additive function
$d(\yv,\cv) = \sum_{i=1}^n f_i(c_i,y_i)$, where the $f_i$'s are some suitable nonnegative functions. This problem is generally solved  (and interesting) for parameters $k$ and bound $D$ on the distance
$d(\cdot,\cdot)$ for which there might be even exponentially many codewords meeting the distance bound, but this does not make the problem necessarily easy.
Right now, there are only very special classes of codes for which there is a satisfying solution to this problem such as for instance convolutional codes, LDGM codes or polar codes.
Interestingly enough, decoded quantum interferometry solves this kind of problem by using the quantum Fourier transform to transform a (classical) solver of the decoding problem in the injective regime into
a solver in the surjective regime. Here by injective regime, we mean that the distance bound is so strong, that for a given $\yv$ there is typically at most one solution of the decoding problem, namely at most a unique $\cv \in \C$ such that $d(\yv,\cv) \leq D$. This corresponds to the more traditional decoding problem appearing in error correction where we are given a noisy codeword $\cv + \ev$, where $\cv \in \C$ and $\ev$ is a ``small'' error vector in $\F_q^n$, where we want a decoding algorithm $\aa_{dec}$ which given as input $\cv + \ev$ is able to recover the close codeword $\cv$. We are in this case typically in the regime where it is only $\cv$ which is close enough to $\cv + \ev$. It turns out that this way of turning a classical algorithm solving the decoding problem in the injective regime into a quantum
algorithm solving the decoding problem in the surjective regime is precisely what Regev has achieved in lattice based cryptography when he reduced quantumly the SIS problem to the LWE problem.

\paragraph{Regev's reduction.} To explain Regev's reduction \cite{R05} in coding theoretic terms, let us bring the notion of {\em dual code}. The dual $\C^\bot$ of a linear code $\C \subset \F_q^n$ is a subspace of
$\F_q^n$ defined as
$$\C^{\bot} = \{\yv \in \F_q^n : \yv \cdot \cv = 0 \textrm{ for each } \cv \in \C\}$$
where $\yv \cdot \cv$ is the canonical inner product of these vectors in $\F_q^n$. We fix a function $f : \F_q^n \rightarrow \mathbb{C}$ with $\norm{f}_2 = 1$ so that $|f|^2$ will serve as a probability distribution of the errors we feed in the decoder. Regev's reduction consists in performing the following operations:
$$
\frac{1}{\sqrt{|\C|}} \sum_{\cv \in \C, \ev \in \F_q^n} f(\ev) \ket{\cv + \ev}\ket{\ev} \xrightarrow{\textrm{Decoder}} \frac{1}{\sqrt{|\C|}} \sum_{\cv \in \C, \ev \in \F_q^n} f(\ev) \ket{\cv + \ev} \xrightarrow{QFT_{\F_q^n}} \frac{q^n}{\sqrt{|\C|}} \sum_{\yv \in \C^\bot} \hf(\yv)\ket{\yv},$$
where $QFT_{\F_q^n}$ denotes the quantum Fourier transform with respect to $\F_q^n$.
The initial state can be easily constructed from a uniform superposition of codewords tensored with a superposition of errors $\frac{1}{\sqrt{|\C|}} \sum_{\cv \in \C} \ket{\cv} \otimes \sum_{\ev \in \F_q^n} f(\ev)\ket{\ev}$
and adding the second register to the first one to create the entangled state $\frac{1}{\sqrt{|\C|}} \sum_{\cv \in \C, \ev \in \F_q^n} f(\ev) \ket{\cv + \ev}\ket{\ev}$.

The idea of Regev's algorithm is to make use of a classical decoder which is an algorithm $\aa_{dec}$ such that $\aa_{dec}(\cv + \ev) = \ev$.
To be able to perform this operation, we are necessarily in the aforementioned injective regime of decoding. Such an algorithm, if it exists, allows us to ``erase" the $\ket{\ev}$ register when applied coherently.
The resulting state $\frac{1}{\sqrt{|\C|}} \sum_{\cv \in \C, \ev \in \F_q^n} f(\ev) \ket{\cv + \ev}=\sum_{\yv \in \F_q^n} \alpha_\yv \ket{\yv}$ is periodic in the sense that its amplitudes $\alpha_\yv$ clearly satisfy $\alpha_{\yv} = \alpha_{\yv+\cv}$ for any $\cv$ in $\C$. This explains why when we apply the
quantum Fourier transform to it, we only get non zero amplitudes on the orthogonal space $\C^\bot$. Measuring the final state will give a dual codeword according to a probability distribution proportional to $|\hf|^2$ restricted on the dual code. If $|f(\ev)|^2$ concentrates its weight on small $\ev$'s, then it turns out that in many examples of interest $|\hf(\yv)|^2$ also concentrates on rather small $\yv$'s. This gives a way to sample
codewords in $\C^\bot$ of rather small weight/norm.

In other words, this reduction shows that if we have access to a decoder $\aa_{dec}$ for the code $\C$ with error distribution $|f|^2$ for our choice of $f$ then we can sample dual codewords according to a distribution proportional to $|\hf|^2$.
This reduction works exactly the same way in $\mathbb{Z}_q^n$ except we apply the Quantum Fourier Transform $QFT_{\mathbb{Z}_q}^n$ instead. \cite{JSW+24} can be viewed as taking in Regev's reduction, codes $\C$ which can be decoded efficiently in the injective regime, namely LDPC codes in the case of the sparse max-XORSAT problem and Reed-Solomon codes in the case of the OPI problem, together with a well chosen function $f$, to have a sampler in the dual code which yields interesting solutions of the optimization problem.

\paragraph{The $\SIS_{\infty}$ and $\ISIS_{\infty}$ problems.}
In lattice based cryptography, Regev's reduction has also been shown to be a candidate for giving quantum algorithms that provide a quantum advantage \cite{CLZ22}.
This concerns the Short Integer Solution with infinity norm, and its inhomogeneous version, which are standard in lattice-based cryptography. These problems can be defined by changing slightly the code alphabet
which becomes $\Z_q$ which will be conveniently be viewed as $\mathbb{Z}_q = \{-\frac{q-1}{2},\dots,\frac{q-1}{2}\}$. For $x \in \mathbb{Z}_q$, we define $|x| = x$ iff. $x \in \{0,\dots,\frac{q-1}{2}\}$ and $|x| = -x$ otherwise. Finally, for $\xv = (x_1,\dots,x_n)$, we define $\norm{\xv}_{\infty} = \max_{i \in \Iint{1}{n}} |x_i|$. In the case $q$ is prime, we work in $\F_q$ and use the same notation as above. 

\begin{problem}[$\SIS_\infty$ - Short Integer Solution with Infinity Norm] $ \ $ \\
	\textbf{Input:} a code $\C$ specified by a parity-check matrix $\Hm \in \F_q^{n \times (n-k)}$\ie \  $\C = \{\xv \in \F_q^n : \Hm \trsp{\xv} = \zero\}$ with prime $q$, an integer $u \in \{0,\dots,\frac{q-1}{2}\}$.\\
	\textbf{Goal:} \ find a non zero $\yv = (y_1,\dots,y_n) \in \C$ such that $\norm{\yv}_\infty \leq u$.
\end{problem}

Many cryptosystems are based on the hardness of $\SIS_{\infty}$ against classical and quantum computers, such as for instance Lyubashevski signature scheme~\cite{Lyu09,Lyu12} or the NIST-standardized DILITHIUM\footnote{ \url{https://pq-crystals.org/dilithium/}}.
A closely related problem is the Inhomogeneous Short Integer Solution with infinity norm.

\begin{problem}[$\ISIS_\infty$ - Inhomogenous Short Integer Solution with Infinity Norm] $ \ $ \\	
	\textbf{Input:} a code $\C$ specified by a parity-check matrix $\Hm \in \F_q^{n \times (n-k)}$ with prime $q$, an integer $u \in \{0,\dots,\frac{q-1}{2}\}$, a random syndrome $\sv \in \F_q^{n-k}$.\\	
	\textbf{Goal:} \ find $\yv \in \F_q^n$ such that $\Hm \trsp{\yv} = \trsp{\sv}$ and $\norm{\yv}_{\infty} \le u$.
	\end{problem}
\cite{CLZ22} gives a polynomial time quantum algorithm for solving the $\SIS_\infty$ problem by considering in Regev's reduction a suitable function $f$ 
such that $\Supp(\hf) \subseteq \{\yv \in \F_q^n : \norm{\yv}_\infty \le u\}$ and by improving Regev's reduction.  The improvement comes from noticing that in Regev's reduction instead of solving the classical
problem asking to recover $\ev$ from the noisy codeword $\cv + \ev$, we may as well recover $\cv$ from the knowledge of the quantum superposition
$\sum_{\ev} f(\ev) \ket{\cv + \ev}$ (this is the so called quantum decoding problem \cite{CT24}).
This can be used to create the superposition $\frac{1}{\sqrt{|\C|}} \sum_{\cv \in \C, \ev \in \F_q^n} f(\ev) \ket{\cv + \ev}$ if we start instead with the superposition
$\frac{1}{\sqrt{|\C|}} \sum_{\cv \in \C, \ev \in \F_q^n} f(\ev) \ket{\cv + \ev}\ket{\cv}$ and use a quantum solver of the quantum decoding problem to erase the second register. There are some
reasons to believe that this quantum decoding problem is easier to solve than the classical decoding problem \cite{CLZ22,CT24}. In \cite{CLZ22} the authors are able to achieve this for random linear codes
in a regime where $k$ is quite small compared to $n$ thanks to a filtering technique and the Arora-Ge algorithm \cite{AG11}. For the regime of parameters considered in \cite{CLZ22}, it turns out that the best known algorithms have subexponential complexity.

It is worthwhile to note that 
Regev's reduction can be easily modified, using the same decoding algorithm in order to construct $\sum_{\yv  \in \vv+\C^\perp} \hf(\yv) \ket{\yv}$ for any $\vv \in \Fq^n$. Indeed, Regev's reduction is slightly modified as follows
$$
\frac{1}{\sqrt{|\C|}} \sum_{\cv \in \C, \ev \in \F_q^n} \chi_\cv(-\vv)f(\ev) \ket{\cv + \ev}\ket{\ev} \xrightarrow{\textrm{Decoder}} \frac{1}{\sqrt{|\C|}} \sum_{\cv \in \C, \ev \in \F_q^n} \chi_\cv(-\vv)f(\ev) \ket{\cv + \ev} \xrightarrow{QFT_{\F_q^n}} \frac{q^n}{\sqrt{|\C|}} \sum_{\yv \in \vv+\C^\perp} \hf(\yv)\ket{\yv}.$$ Here $\chi_\cv$ is the character of $\F_q^n$ associated to $\cv$ (see \S \ref{ss:fourier}), the Fourier transform over $\Fqn$ being defined by $\hat{f}(\yv) \eqdef \frac{1}{\sqrt{q^n}} \sum_{\xv \in \Fqn}
\chi_{\yv}(\xv) f(\xv)$. The fact that the quantum Fourier transform yields a state of the form  $\frac{q^n}{\sqrt{|\C|}} \sum_{\yv \in \vv+\C^\perp} \hf(\yv)\ket{\yv}$ follows on the spot from Proposition \ref{proposition:periodic} in \S \ref{ss:fourier}. This means we can solve the (I)$\SIS_{\infty}$ problem for $\C^{\bot}$ as long as we have a decoder for $\C$ w.r.t. function $f$.

\paragraph{Improving Regev's reduction by taking better decoders.} 

In order to have the most of Regev's reduction, it makes sense to move to decoders which have a better decoding capacity. This is well illustrated in \cite{DRT23} where Regev's reduction is analyzed in the context of
random linear codes and for producing codewords of low Hamming weight. There it is proved that the higher the Hamming weight corrected by the decoding algorithm for $\C$, the smaller the weight
of the dual codewords which are  produced become. In order to have a useful reduction, the authors of \cite{DRT23} had for many parameters to go beyond the Hamming weight $t$ for which there is always a unique solution to the decoding problem.
Another example could be the OPI problem considered in \cite{JSW+24}. As explained above, the authors solve this problem through Regev's reduction applied to Reed-Solomon codes. These codes can be readily decoded by the Berlekamp-Welch decoder
which decode all errors up to half the minimum distance of the code. This is precisely the largest Hamming weight for which unique decoding is ensured. Potentially we could improve the OPI quantum algorithm of \cite{JSW+24} by using instead
of the Berlekamp-Welch decoder an algorithm which decodes more errors, such as the Guruswami-Sudan list decoder \cite{GS98} or the Koetter-Vardy decoding algorithm \cite{KV03}. However, this means that we have to use a decoding algorithm beyond the limit where we can ensure that it always recovers  from $\cv+\ev$ the right error $\ev$. It would be particularly nice to be able to use the Koetter-Vardy decoding algorithm, since it is able to use {\em soft information}. To explain
what we mean by this consider a random codeword $\cv = (c_1,\dots,c_n)$ and assume that we  add an error on each coordinate according to some probability function $p : \F_q \rightarrow \mathbb{R}_+$. The efficiency of this decoder will not only depend on the amount of errors ({\ie}, $n(1 - p(0))$ on average) but actually on the error distribution in general and this could be helpful to improve the amount of errors that the decoder can correct. Such decoders are called {\em soft decoders}.
However, there is an issue: while the Berlekamp-Welch decoder always outputs the correct value, the other decoders work in parameters where there are necessarily some decoding errors. Small errors may not seem very problematic for the overall algorithm but it can actually be devastating as we now recall.





\paragraph{The issue with errors in the decoder.}
One important technical difficulty when dealing with this family of algorithms is that small errors in the decoding algorithm can lead to large errors when measuring the dual codeword. Say we have a function $f : \F_q^n \rightarrow \mathbb{C}$ with $\norm{f}_2 = 1$. Now assume there is a decoding algorithm $\aa_{dec}$ and a set $\GOOD \subseteq \F_q^n$ such that $\forall \cv \in \C$, $\forall \ev \in \GOOD$,  $\aa_{dec}(\cv + \ev) = \ev$. In the ideal setting, we want to create $\ket{\phi_{\Ideal}}$ while we actually create the state $\ket{\phi_{\Real}}$ where:
$$
\ket{\phi_{\Ideal}}  = \frac{1}{\sqrt{Z}} \sum_{\substack{\cv \in \C \\ \ev \in \F_q^n}} f(\ev) \ket{\cv + \ev} \quad ; \quad 
\ket{\phi_{\Real}}  = \frac{1}{\sqrt{Z'}} \sum_{\substack{\cv \in \C \\ \ev \in \GOOD}} f(\ev) \ket{\cv + \ev}
$$
where $Z,Z'$ are normalizing factors. Now assume our decoder succeeds with probability $1-\eps$ which means $\sum_{\ev \in \GOOD} |f(\ev)|^2 = 1-\eps$. It is possible to construct a function $f$ and a decoder where $\eps = o(1)$ but $|\braket{\phi_{\Ideal}}{\phi_{\Real}}| = \negl(n)$. Some examples of this phenomenon were provided in~\cite{CT24}, and this means that there is no hope of a generic and unconditional reduction between decoding $\C$ according to $|f|^2$ and sampling dual codewords according to $|\hf|^2$ in the case we have errors in the decoder. This is a pity, since as explained before in order to improve the reduction we are naturally lead to use decoders which can not decode all errors we are interested in and dealing with this kind of issue is really delicate due to to the complicated shape of the errors which can not be decoded by a given code/decoder pair. The example of the previous Reed-Solomon code/Guruswami-Sudan algorithm is an example of this kind.

Another closely related setting is the inhomogeneous setting. Here, we have a random dual syndrome $\sv$ and we want a reduction theorem that says that  if we can decode a code $\C$ according to $|f|^2$ in polynomial time then we can sample words from the dual coset $\C_\sv^{\bot} = \{\yv \in \F_q^n : \Hm^\bot \trsp{\yv} = \trsp{\sv}\}$ according to $|\hf|^2$ where $\Hm^\bot$ is a parity matrix of $\C^\bot$. The previous case corresponds to the case $\sv = 0$. The computational complexity of the problem for a random $\sv$ is most often at least as hard as the homogeneous setting. Moreover, we do not know whether errors in the decoder can cause the reduction to collapse.

\COMMENT{In the inhomogeneous setting, we have a dual syndrome $\sv \in \F_q^{k}$ and let $\vv \in \F_q^n$ be any vector such that $\Hm^{\bot} \vv = \sv$. The ideal and real states become (presented here for the case of a prime $q$)

$$
\ket{\phi_{\Ideal}^{\vv}}  = \frac{1}{\sqrt{Z}} \sum_{\substack{\cv \in \C \\ \ev \in \F_q^n}} F(\ev) \omega_q^{\vv \cdot \cv} \ket{\cv + \ev} \quad ; \quad 
\ket{\phi_{\Real}^{\vv}}  = \frac{1}{\sqrt{Z'}} \sum_{\substack{\cv \in \C \\ \ev \in \GOOD}} F(\ev) \omega_q^{\vv \cdot \cv}  \ket{\cv + \ev}
$$}

\subsection{Overview of our contributions}
We have two main contributions:
\begin{enumerate}
	\item We first show a that in the inhomogeneous setting, the reduction always works even with errors in the decoder. This result is generic and captures the full power of quantum algorithms based on Regev's reduction. 
	\item We apply the above result for the case of Reed-Solomon codes using the Koetter and Vardy soft decoder. We provide powerful quantum polynomial algorithms for the Optimal Polynomial Interpolation problem. In the particular case of $\ISIS_{\infty}$ on Reed-Solomon codes, we show for example how to solve the problem for $u \approx \frac{q}{4}$ in quantum polynomial time for $k = \frac{2n}{3}$ while the decoded quantum interferometry algorithm cannot solve this problem except for $k \ge n(1 - o(1))$. This provides improvements on the problems that can be proposed for quantum advantage in this setting.
\end{enumerate}

\subsubsection{Generic reduction theorem for the inhomogeneous setting}\label{Section:IntroReduction}
Our first result is to show that indeed a generic reduction theorem holds in the inhomogeneous setting. This is the first time there is a unconditional reduction with errors in the decoder.
This can be considered as an important step for using Regev's reduction beyond the unique decoding regime. As explained before, this is indeed important to improve the quality of the dual codewords we find through this reduction.
In order to present our reduction theorem, we first define the problems that will be involved.


\begin{problem}[Syndrome Decoding - $\SD(\Hm,p)$] $ \ $ \\	
	\textbf{Given :} $(\Hm,\Hm \trsp{\ev})$ where $\Hm \in \F_q^{(n-k)\times n}$ and $\ev$ is distributed according to some  probability function $p : \F_q^n \rightarrow \mathbb{R}_+$.\\	
	\textbf{Goal :} Find $\ev$.
\end{problem}

This is basically an equivalent form of the decoding problem we mentioned in the introduction which will be more convenient here. The fact that this is an equivalent form of the decoding problem can be verified as follows. From the knowledge of a noisy codeword $\cv + \ev$ and a parity-check matrix $\Hm$ of the code $\C$ we want to decode, meaning that $\C = \{\cv \in \F_q^n: \Hm \trsp{\cv} = \zero\}$, we can compute $\Hm \trsp{(\cv+\ev)} = \Hm \trsp{\ev}$ and use this knowledge to recover $\ev$ by solving the syndrome decoding problem. 


\begin{definition}
	For a probability function $p : \F_q^n \rightarrow \mathbb{R}_+$, a parity-check matrix $\Hm \in \F_q^{(n-k)\times n}$ and a syndrome $\sv \in \F_q^{n-k}$, we define the probability function $u_p^{\Hm,\sv}$ as follows
	\begin{align*}
		u_p^{\Hm,\sv}(\xv) = \left\{
		\begin{array}{cl} \frac{p(\xv)}{\sum_{\xv : \Hm \trsp{\xv} = \trsp{\sv}} p(\xv)} & \textrm{ if } \Hm \trsp{\xv} = \trsp{\sv} \\
			0 & \textrm{ otherwise }
		\end{array}
		\right. 
	\end{align*}
	$u_p^{\Hm,\sv}$ corresponds to the probability function $p$ restricted to the coset $\{\xv \in \F_q^n : \Hm \trsp{\xv} = \sv\}$. In the case ${\sum_{\xv : \Hm \trsp{\xv} = \trsp{\sv}} p(\xv)} = 0$, we just define the above function to be the null function (in which case we say that sampling from this function always fails).
\end{definition}

Basically with the help of an algorithm solving the syndrome decoding problem and Regev's reduction we wish to solve the following coset sampling problem.

\begin{problem}[Coset Sampling - $\CSP(\Hm,p)$] $ \ $
	
	\textbf{Given :} $(\Hm,\sv)$ where $\Hm \in \F_q^{(n-k)\times n}$  and $\sv \Unif \F_q^{n-k}$
	
	\textbf{Goal :} sample from the probability function
	$
	u_p^{\Hm,\sv}.
	$ 
\end{problem}

Our reduction relates the syndrome decoding problem on a parity matrix $\Hm$ of a code $\C$ and the coset sampling problem on a parity $\Hm^{\bot}$ of the dual code $\C^{\bot}$. It can informally be described as follows

\begin{theorem}[Informal]\label{Theorem:1Intro}
For a function $f : \F_q^n \rightarrow \mathbb{C}$, if we have access to an algorithm that efficiently solves $\SD(\Hm,|f|^2)$ with some error $P_{dec} \ge \frac{1}{\poly(n)}$ then we can construct an efficient quantum algorithm for  $\CSP(\Hm^{\bot},|\hf|^2)$ that will sample outputs with a probability function at least $\frac{1}{\poly(n)}$ close to  $u_{|\hf|^2}^{\Hm,\sv}$ on average on $\sv$.
\end{theorem}

Roughly speaking, what this reduction means, is that as soon as we have an algorithm solving the decoding problem with probability at least $1/\poly$ when the error is distributed according to the  distribution $|f|^2$, then we can efficiently sample in the coset $\{\xv \in \F_q^n: \Hm \trsp{\xv} = \trsp{\sv}\}$ roughly as the distribution $|\hf|^2$ conditioned that we are in this coset for most of the $\sv$'s. In other words, this means that we have a way to solve the inhomogeneous decoding/sampling problem for most of entries $\sv$.

\paragraph{Comparison with existing algorithms.} It is natural to compare this theorem with existing similar theorems. Indeed, all the works using quantum algorithms based on Regev's reduction must provide some variant of the above theorem to work.

First, most results deal with the homogeneous case, meaning that $\sv = \zerov$ in the sampling problem. This case works fine when $P_{dec} = 1$ (in which case the faithfulness\footnote{See Definition \ref{def:faithfulness} for a definition of faithfulness.} is $1$) but
we know we cannot achieve with Regev's reduction an equivalent of Theorem~\ref{Theorem:1Intro}. Indeed, we know examples where $P_{dec} = 1 - o(1)$ but the resulting algorithm has faithfulness $\negl(n)$ (see~\cite{CT24} for some examples).  In~\cite{CLZ22}, the authors are in a regime where $1-P_{dec} \ll q^{-k}$ in which case the reduction works similarly to the perfect setting. They obtain such small errors in the decoder by considering the setting where $n$ is superlinear in $k$. In~\cite{YZ24}, the authors manage to prove that their real and ideal setting are close by using the randomness of the random oracle inherent to their problem. In the adaptation of Regev's reduction for codes, the authors of~\cite{DRT24} overcome the impossibility result by restricting themselves to real non-negative functions $f$ and also random matrices $\Hm \in \F_q^n$. In~\cite{DFS24}, the authors perform oblivious sampling using Regev's reduction. They use a quantum decoder which is unambiguous {\ie} it always outputs the correct value or aborts. In our work, we are in none of the situations above so we cannot use these results.

Finally, the case the closest to our setting is the work of~\cite{JSW+24}. They also consider the inhomogeneous setting. They manage to prove something equivalent to Theorem~\ref{Theorem:1Intro}. In their setting, they consider random matrices $\Hm$ according to some specific distribution and need a heuristic on the weight distribution of the underlying codes in order to prove that the reduction works. As a direct consequence, Theorem~\ref{Theorem:1Intro} implies that this claim of~\cite{JSW+24} can be done without any heuristic (Conjecture $7.1$ from~\cite{JSW+24}). 

\subsubsection{Using the KV decoder for solving the $\SAINTERPOL$ problem}

The Koetter-Vardy decoder will allow us to solve a more constrained problem than the OPI problem in a regime of parameters that the \cite{JSW+24} is unable to handle. In a nutshell, instead of trying to obtain a polynomial meeting a large fraction of  the  intersection constraints in the OPI problem we want to satisfy {\em all of them}. This can be rephrased as the following   $\SAINTERPOL$ problem.
\begin{problem}[$\SAINTERPOL$ problem] $ $ \\
	\textbf{Given: } a uniformly random  $\yv=(y_\alpha)_{\alpha \in \fq} \in (\fq)^q$, an integer $k$ in $\Iint{1}{q}$, a fixed subset 
	$\cS$ of $\fq$ \\
	\textbf{Goal:}  find a polynomial $P(X) \in \fq[X]$ of degree $<k$ such that $y_\alpha - P(\alpha) \in \cS$ for 
	any $\alpha \in \fq$.
\end{problem}

Notice here that we consider here the full support setting, {\ie}, $n = q$. This ensures that the dual of a Reed-Solomon code is a Reed-Solomon code and simplifies a little bit the statements and the discussion, but strictly speaking this is not needed.
In a sense, this is a more natural problem than the OPI problem. An important case is when $q$ is prime and $\cS=\Iint{-u}{u}$, in which case, this is actually a $\ISIS_{\infty}$ problem for Reed Solomon codes which is defined more formally in Section~\ref{Section:RSProblem}. We call it the $\RSISIS_\infty$ problem. Informally, we have a vector $\sv$ in $\F_q^{q-k}$ and want to find
$\ev$ in $\F_q^q$ such that we have at the same time $\Hm \trsp{\ev} = \trsp{\sv}$ and $\norm{\ev}_\infty \leq u$. Here $\Hm$ is parity-check matrix of a Reed-Solomon code over $\fq$ of full length $q$.

Using our reduction theorem combined with the Koetter and Vardy soft decoder, we manage to show the following 

\begin{theorem}\label{Theorem:ISISfIntro}
	Consider any function $f : \F_q \rightarrow \mathbb{C}$ st. $\Supp(\hh) \in \cS$ and $\norm{f}_2 = 1$. There exists a quantum polynomial time algorithm for the $\SAINTERPOL$ problem defined above as long as $\frac{k}{q} > 1 -  \sum_{\alpha \in \F_q} |f(\alpha)|^4$.
\end{theorem}

Remark that the above theorem, we consider functions $f : \F_q \rightarrow \mathbb{C}$ and the corresponding error function is actually $f^{\otimes n}$. In the theorem above, the efficiency of our algorithm depends actually on the choice of the function $f$ and there are many candidates satisfying $\Supp(\hf) \in \cS$. In the case we pick $\hf = \frac{1}{\sqrt{|\cS|}}\one_{\cS}$, we manage to prove the following, first for the case of a randomly chosen $\cS$.

\begin{theorem}\label{Proposition:AlgoSIntro}
	There exists a quantum polynomial time algorithm that solves $\SAINTERPOL$  with parameters $q,k,S$ as long as $\frac{k}{q} < 1 - \rho^2$ where $\rho = \frac{s}{q}$, with high probability over a randomly chosen $S \subseteq \F_q$ of size $s$.
\end{theorem}

In the specific case of $\ISIS_{\infty}$, meaning that we take $\cS = \Iint{-u}{u}$, we obtain the following stronger result.

\begin{theorem}
	There exists a quantum algorithm that solves $\RSISIS_\infty$ with parameters $q,k,u$ as long as $k \ge V(\rho)q$ where $\rho = \frac{|S|}{q} = \frac{2u+1}{q}$ and 
	\begin{align*}
		V(\rho) = \left\{ 
		\begin{array}{ll}
			1 - \frac{2\rho}{3} & \textrm{for } \rho \le \frac{1}{2} \\
			5 - \frac{10\rho}{3} - \frac{2}{\rho} + \frac{1}{3\rho^2} & \textrm{for } \rho \ge \frac{1}{2}
		\end{array} \right.
	\end{align*}
\end{theorem}

\paragraph{Comparison with existing work.} We compare ourselves with the Decoded Quantum Interferometry algorithm of~\cite{JSW+24} which gives the following

\begin{proposition}[\cite{JSW+24}]
	There is a quantum polynomial time algorithm that solves $\SAINTERPOL$ with parameters $q,k,S$ as long as $\frac{k}{q} \ge \min(1,2-2\rho)$ where $\rho = \frac{|S|}{q}$.
\end{proposition}

\begin{figure}[!ht]
	\begin{center}
		\includegraphics[width = 10cm]{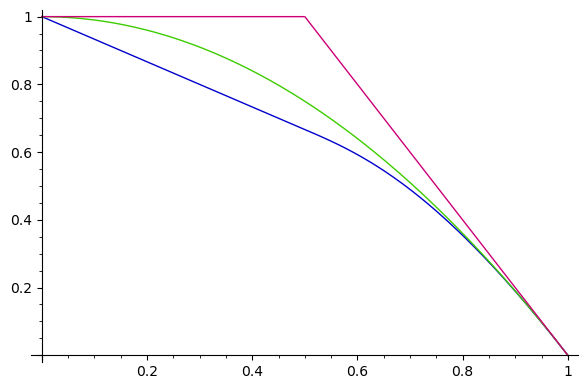} \end{center}
	\caption{Minimum admissible degree ratio $R = \frac{k}{n}$ as a function of $\rho = \frac{|S|}{q}$. The (top) purple line corresponds to the \emph{Decoded Quantum Interferometry} algorithm for $\SAINTERPOL$. The (middle) green line corresponds to our algorithm for $\SAINTERPOL$ with random $S$ and the (bottom) blue line corresponds to  our algorithm with $S = \Iint{-u}{u}$.}
	\label{Figure:QA}
\end{figure}

\noindent The comparison between the three algorithms is depicted in Figure~\ref{Figure:QA}.

\paragraph{How to read this plot.}
We can read the above plot as follows. Consider the case $S \approx \frac{q}{2}$, the Decoded Quantum Interferometry will find a polynomial with degree $k \approx q$, no matter the structure of $S$. This is dangerously close to the case when the problem becomes trivial. With our algorithm, we get improvements depending on the structure of $S$. If $S$ is chosen randomly from a subset of size approximately $\frac{q}{2}$, then we can find a polynomial satisfying the constraints with a degree $k \le \frac{3q}{4}$. Finally, if we additionally have the requirement that $S = \Iint{-u}{u}$ (with still $|S| \approx \frac{q}{2})$, then our algorithm finds a polynomial satisfying the constraints with a degree $k \le \frac{2q}{3}$.

\paragraph{Other results.} Even if the uniform function $\hf$ on $\cS$ is appealing, nothing says that this will be the optimal function in Theorem~\ref{Theorem:ISISfIntro}. What we show is that the uniform function is actually in general not optimal. For the special case of $u=1$ in $\SIS_{\infty}$, we provide the optimal function that performs better for any $q \ge 5$. 
We also provide a numerical analysis for $q = 1009$ and all values of $u \in \Iint{0}{(q-1)/2}$ for some non uniform function $\hf$ that slightly outperforms the uniform function. Finally, we present the best known classical algorithm for this problem and compare with our quantum algorithm. 

\section{Preliminaries}
\subsection{Notation}\label{ss:notation}
\paragraph{Sets, finite field.}
The finite field with $q$ elements is denoted by $\Fq$. 
$\Zq$ denotes the ring of integers modulo $q$. We will represent the elements of $\Fq$ when $q$ is an odd prime as
$\F_q = \{-\frac{q-1}{2},\dots,\frac{q-1}{2}\}$ and also use in this case the notation $ \omega_q = e^{\frac{2i\pi}{q}}$.
The cardinality of a finite set $\Ec$ is denoted by $|\Ec|$.
The set of integers $\{a,a+1,\cdots,b\}$ between the integers $a$ and $b$ is denoted by $\Iint{a}{b}$. For a positive integer $n$, $[n]$ denotes 
$\Iint{1}{n}$. 
\paragraph{Vector and matrices.}
Vectors are row vectors as is standard in the coding community and $\trsp{\xv}$ denotes the transpose of a vector or a matrix. In particular, vectors will always be denoted by bold small letters and matrices with bold capital letters. 

\paragraph{Functions.} Functions defined on $\Fq$ will be denoted by small letters, such as $f$ or $g$, whereas functions defined on $\Fqn$ we be denoted by capital letters such as $F$ and $G$. The tensor product $f \otimes g$ of two functions $f:\fq \rightarrow \Comp$ and $g:\fq \rightarrow \Comp$ is a function from $\F_q^2$ to $\Comp$ defined as $(f \otimes g) (x_1,x_2) = f(x_1)g(x_2)$ for all $(x_1,x_2)$ in $\fq \times \fq$. For two functions $f,g : \F_q \rightarrow \mathbb{C}$, their convolution is defined as $(f \star g) (x) = \sum_{y \in \F_q} f(y)g(x-y)$. For a function $f:\fq \rightarrow \Comp$, its $n$-th tensor power $f^{\otimes n}$ is a function $f^{\otimes n}: \Fqn \rightarrow \Comp$ defined by 
$f^{\otimes n}(x_1,\cdots,x_n) = f(x_1)\cdots f(x_n)$ for all $(x_1,\cdots,x_n) \in \Fqn$, or equivalently $f^{\otimes n}=\underbrace{f \otimes \cdots \otimes f}_{n \text{ times}}$.
\paragraph{Probability.}
	A (discrete) probability (mass) function $p$ is a non-negative real function defined on a discrete set $\Xc$ satisfying $\sum_{x \in \Xc} p(x) = 1$. For a finite set $\Xc$ we denote by $X \Unif \Xc$ a random variable distributed uniformly on $\Xc$. For two probability functions $p,r : \Xc \rightarrow \mathbb{R}_+$, we define their trace distance $\Delta(p,r) = \frac{1}{2} \sum_{x \in \Xc} |p(x) - r(x)|$. We also extend this definition to quantum pure states and define $\Delta{(\ket{\psi},\ket{\phi})} = \sqrt{1 - |\braket{\psi}{\phi}|^2}$.

\subsection{The classical and quantum Fourier transform on $\Fqn$}
\label{ss:fourier} b
In this article, we will use the quantum Fourier transform on $\Fqn$.
\paragraph{\bf Definition and basic properties.}
It is based on the characters of the group $(\Fqn,+)$ which are defined as follows (for more details see \cite[Chap 5, \S 1]{LN97}, in particular a description of the characters in terms of the trace function is given in  \cite[Ch. 5, \S 1, Th. 5.7]{LN97}).
\begin{definition}
	Fix $q = p^s$ for a prime integer $p$ and an integer $s \ge 1$. The characters of $\F_q$ are the functions $\carf{y} : \F_q \rightarrow \mathbb{C}$ indexed by elements $y \in \F_q$ defined as follows
	\begin{eqnarray*}
		\car{y}{x} & \eqdef & e^{\frac{2i \pi \tr(x \cdot y)}{p}}, \quad \text{with} \\
		\tr(a) & \eqdef & a + a^p + a^{p^2} + \dots + a^{p^{s-1}}.
	\end{eqnarray*}
	where the product $x \cdot y$ corresponds to the product of elements in $\F_q$. We extend the definition to vectors $\xv,\yv \in \F_q^n$ as follows:
	$$ \car{\yv}{\xv} \eqdef \Pi_{i = 1}^n \car{y_i}{x_i}.$$
\end{definition}
When $q$ is prime, we have $\car{y}{x} = e^{\frac{2i\pi xy}{q}}=\omega_q^x$. In the case where $q$ is not prime, the above definition is not necessarily easy to handle for computations. Fortunately, characters have many desirable properties that we can use for our calculations.
\begin{proposition}\label{Proposition;Characters}
	The characters $\carf{\yv} : \Fqn \rightarrow \mathbb{C}$ have the following properties
	\begin{enumerate}\setlength\itemsep{-0.2em}
		\item (Group Homomorphism). $\forall \yv \in \Fqn$, $\carf{\yv}$ is a group homomorphism from $(\Fqn,+)$ to $(\mathbb{C},\cdot)$ meaning that $\forall \xv, \xv' \in \Fqn$, $\car{\yv}{\xv + \xv'} = \car{\yv}{\xv} \cdot \car{\yv}{\xv'}$.
		\item \label{eq:symmetry} (Symmetry). $\forall \xv, \yv \in \Fqn, \ \car{\yv}{\xv} = \car{\xv}{\yv}$
		\item \label{eq:orthogonality} (Orthogonality of characters). The characters are orthogonal functions meaning that $\forall \xv, \xv' \in \Fqn$, $\sum_{\yv \in \Fqn} \car{\yv}{\xv}\overline{\car{\yv}{\xv'}} = q^n \delta_{\xv,\xv'}$. In particular $\sum_{\yv \in \Fqn} |\car{\yv}{\xv}|^2 = q$ and $\forall \xv \in \Fqn\setminus\{0\}, \ \sum_{\yv \in \Fqn} \car{\yv}{\xv} = 0$.
	\end{enumerate}
\end{proposition}

Notice that these imply some other properties on characters. For instance $\chi_\yv(\mathbf{0}) = 1$ or $|\car{\yv}{\xv})| = 1$ for any $\xv,\yv \in \Fqn$.
The orthogonality of characters, allows to define a unitary transform which is is nothing but  the classical or the quantum Fourier transform on $\Fqn$.

\begin{definition}
	For a function $ F : \Fqn \rightarrow \C$, we define the (classical) Fourier transform $\hat{F}$ as
	$$
	\FT{F}(\xv) = \frac{1}{\sqrt{q^n}} \sum_{\yv \in \Fqn} \car{\xv}{\yv} F(\yv).
	$$
	The quantum Fourier transform $\QFTt$ on $\Fqn$ is the quantum unitary satisfying $\forall \xv \in \Fqn$, 
	\begin{eqnarray*}
		\QFTt \ket{\xv} &= &\frac{1}{\sqrt{q^n}} \sum_{\yv \in \Fqn} \chi_{\xv}(\yv) \ket{\yv}.
	\end{eqnarray*}
	We will also write $\ket{\QFT{\psi}} \eqdef \QFTt \ket{\psi}$.
\end{definition}
Note that when $\ket{\psi} = \sum_{\xv \in \Fqn} F(\xv) \ket{\xv}$ we have
\begin{equation*}
\ket{\QFT{\psi}} = \sum_{\xv \in \Fqn} \FT{F}(\xv) \ket{\xv}.
\end{equation*}
The Fourier transform can also be viewed as expressing the coefficients of a state  in the Fourier basis 
$\left\{\ket{\QFT{\xv}},\xv \in \Fqn \right\}$ as shown by

\begin{fact}\label{fact:inverse}
	Let $\ket{\psi}= \sum_{\yv \in \Fqn} F(\yv) \ket{\yv}$, then
	$$
	\ket{\psi} = \sum_{\xv \in \Fqn} \FT{F}(-\xv) \ket{\QFT{\xv}}.
	$$
\end{fact}
This follows on the spot from the fact that if $\ket{\psi} = \sum_{\xv \in \Fqn} c_{\xv} \ket{\QFT{\xv}}$, then 
$$
c_{\xv} = \braket{\QFT{\xv}}{\psi}= \frac{1}{\sqrt{q^n}} \sum_{\yv \in \Fqn} \overline{\chi_{\xv}(y)} F(y) = \FT{F}(-\xv).
$$


Note that the main properties of the Fourier transform follow from the fact that the characters are the common eigenbasis of all shift operators (and therefore all convolution operators). 
\paragraph{\bf Applying the quantum Fourier transform on periodic states.}
Regev's reduction applies to states which are periodic. They are  of the form $\frac{1}{\sqrt{Z}} \sum_{\cv \in \C}\sum_{\ev \in \F_q^n} F(\ev) \ket{\cv + \ev}$ where 
$Z$ is some normalizing constant, $\C$ some linear code of length $n$ over $\Fq$ and $F$ some function from $\F_q^n$ to $\Comp$. This state can be written as $\frac{1}{\sqrt{Z}} \sum_{\xv \in \F_q^n} G(\xv) \ket{\xv}$ where $G(\xv) = \sum_{\cv \in \C} F(\xv-\cv)$. We clearly have in this 
case $G(\xv +\cv)=G(\xv)$ for any $\xv \in \F_q^n$ and any $\cv \in \C$. The Fourier transform of such periodic states has support in this case $\C^\perp$. In our case, this holds with a slight tweak as shown by 

\begin{proposition}\label{proposition:periodic}
Consider a function $F : \F_q^n \mapsto \Comp$.  We have for all linear codes $\C \subseteq \F_q^n$ and all 
$\vv \in \Fqn$:
$$
\QFTt \left( \sum_{\cv \in \C}\sum_{\ev \in \F_q^n} F(\ev) \chi_\cv(\vv)\ket{\cv + \ev} \right) = |\C| \sum_{\yv \in -\vv + \C^\perp} \QFT{F}(\yv) \ket{\yv}.
$$
\end{proposition}
\begin{proof}
The proposition follows from the following computation
\begin{eqnarray}
\QFTt \left(  \sum_{\cv \in \C}\sum_{\ev \in \F_q^n} F(\ev)  \chi_\cv(\vv) \ket{\cv + \ev} \right) &=&  \sum_{\ev \in \F_q^n} F(\ev)   \sum_{\cv \in \C} \chi_\cv(\vv) \sum_{\yv \in \F_q^n} \chi_\yv(\cv+\ev) \ket{\yv} \nonumber \\
& = &  \sum_{\ev \in \F_q^n} F(\ev)    \sum_{\yv \in \F_q^n}  \chi_\yv(\ev) \ket{\yv}\sum_{\cv \in \C} \chi_\cv(\vv) \chi_{\yv}(\cv) \nonumber \\
& = &  \sum_{\ev \in \F_q^n} F(\ev)    \sum_{\yv \in \F_q^n}  \chi_\yv(\ev) \ket{\yv}\sum_{\cv \in \C} \chi_{\yv+\vv}(\cv) \nonumber \\
& = & |\C| \sum_{\ev \in \F_q^n}  F(\ev) \sum_{\yv \in -\vv+\C^\perp} \chi_\yv(\ev)  \ket{\yv} \label{eq:periodic}\\
& = & |\C| \sum_{\yv \in -\vv+\C^\perp} \ket{\yv} \sum_{\ev \in \F_q^n}  F(\ev) \chi_\yv(\ev)   \nonumber \\
& =&  |\C| \sum_{\yv \in -\vv + \C^\perp} \QFT{F}(\yv) \ket{\yv} \nonumber 
\end{eqnarray}
where \eqref{eq:periodic} follows from a slight generalization of \eqref{eq:orthogonality} of Proposition \ref{Proposition;Characters}, namely that
\begin{eqnarray*}
\sum_{\cv \in \C} \chi_{\yv+\vv}(\cv) & = & |C| \;\;\text{if $\yv+\vv \in \C^\perp$}\\
& = & 0 \;\; \text{otherwise,}
\end{eqnarray*}
which follows by a similar reasoning by noticing that $\C^\perp$ can be viewed as the set of trivial characters acting on $\C$:
$$\{ \yv \in \F_q^n: \chi_{\yv}(\cv)=1,\;\forall \cv \in \C\}= \C^\perp.$$
\end{proof}

\subsection{Coding theory}

\begin{definition}
	A $q$-ary linear code $\C$ of dimension $k$ and length is a linear subset of $\F_q^n$. It can be specified by a generating matrix $\Gm \in \F_q^{k \times n}$, in which case 
	$$ \C = \{\xv \Gm : \xv \in \F_q^k\}.$$
	It can also be specified by a parity matrix $\Hm \in \F_q^{(n-k)\times n}$ in which case
	$$ \C = \{\cv \in \F_q^n : \Hm \trsp{\cv} = \trsp{\zerov}\}.$$
\end{definition}

\begin{definition}
	For a code $\C$ specified by a parity matrix $\Hm \in \F_q^{(n-k)\times n}$ and $\sv \in \fq^{n-k}$, we define $\C_\sv = \{\yv \in \F_q^n : \Hm \trsp{\yv} = \trsp{\sv}\}$. In particular, $\C_{\zero} =\C$.
\end{definition}

\begin{definition}\label{Definition:Decoder}
  For a code $\C$, a decoding algorithm $\aa_{dec}$ takes as input a noisy codeword $\cv + \ev \in \F_q^n$ and tries to recover $\ev$ (hence $\cv$).
  We assume that it always outputs an element in $\Fqn$. Its success probability for an error probability function $p : \F_q^n \rightarrow \mathbb{R}_+$ is given by
	$$ P_{dec} = \Pr_{\cv \in \C,\ev \Unif p} \left[\aa_{dec}(\cv + \ev) = \ev\right].$$
\end{definition}

An important class of decoding algorithms consist, when given a parity-check matrix $\Hm \in \fq^{(n-k)\times n}$ of the code and $\yv \in \F_q^n$ the word which is to be decoded, in\\
(i) computing $\sv$ such that $\trsp{\sv} = \Hm \trsp{\yv}$,\\
(ii) outputting a fixed element $\ev$ in $\F_q^n$ of syndrome $\sv$ \ie\ such that $\Hm \trsp{\ev} = \trsp{\sv}$. In other words a fixed element is chosen in $\C_\sv$ and output by the algorithm.

These algorithms are called syndrome decoders. Equivalently they are characterized by the following property

\begin{definition}[syndrome decoder and decoding set]
  For a code $\C$ specified by its parity-check matrix $\Hm$, a decoding algorithm $\aa_{dec}$ is called a syndrome decoder if it outputs for an element $\yv \in \fq^n$ a fixed element in
  $\C_{\Hm \trsp{\yv}}$. The element which is chosen in $\C_\sv$ is denoted by $\ev_\sv$. The set $\Dc = \{\ev_\sv, \sv \in \fq^{n-k}\}$ is called the decoding set of the syndrome decoder.
\end{definition}

Note that by definition the $\ev_\sv$'s are precisely the errors which we can correct by our syndrome decoder. Indeed, for any $\sv \in \F_q^{n-k}$ and any $\cv \in \C$, the syndrome decoder outputs
by definition $\ev_\sv$ when being fed with $\cv + \ev_\sv$ and never outputs $\ev$ when being given $\cv + \ev$ for $\ev$ which does not belong to $\Dc$. Another important property of the syndrome decoder is
that its answer does not depend on the codeword $\cv$ when being given a noisy codeword $\cv + \ev$: we have $\aa_{dec}(\xv)=\aa_{dec}(\yv)$ iff. $\xv - \yv \in \C$.

\subsection{Reed-Solomon codes and decoding algorithms for these codes}

An important family of codes which have an efficient decoding algorithm is given by the family of Reed-Solomon codes. Since we will also be interested in decoding their dual, it will be convenient to enlarge a slightly bit the family, so that it is at the same invariant by taking the dual and so that we can also decode any member of the family with the same (efficient) decoding algorithm. Generalized Reed-Solomon codes are the right object here.

\begin{definition}[(Generalized) Reed-Solomon codes]
  Let $\xv=(x_1,\dots,x_n)\in\F_q^n$ be a vector of pairwise distinct entries and $\yv=(y_1,\dots,y_n)\in\F_q^n$ a vector of nonzero entries. The \textit{generalized Reed-Solomon (GRS) code} $\GRS{k}{\xv}{\yv}$ with \textit{support} $\xv$, \textit{multiplier} $\yv$ and dimension $k$ is
  $$
  \GRS{r}{\xv}{\yv}\eqdef\{(y_1 P(x_1),\dots,y_n P(x_n)) \mid P \in \F_q[z], \deg P < k\}
  $$
  A Reed-Solomon code $\RS{k}{\xv}$ of dimension $k$ and support $\xv$ is nothing but a GRS code
  with multiplier the constant $1$ vector, \ie 
  $$\RS{k}{\xv}\eqdef \GRS{k}{\xv}{\one}=
  \{(P(x_1),\dots,P(x_n)) \mid P \in \F_q[z], \deg P < k\}.$$
  When $\xv=(x_\alpha)_{\alpha \in \Fq}$ is of length $q$ and its entries is the whole finite field $\fq$, then we simply write $\fsRS{k}$ and say it is the full support Reed-Solomon code of dimension $k$.
  \end{definition}

The dual of a GRS code is also a GRS code as shown by
\begin{proposition} \cite[Theorem~4, p.~304]{MS86}\label{pr:dual_GRS} 
  Let $\GRS{k}{\xv}{\yv}$ be a GRS code of length $n$. Its dual is also a GRS code. In particular
  $$
  \GRS{k}{\xv}{\yv}^\perp=\GRS{n-k}{\xv}{\yv^\perp},
  $$
  where 
  $$
  \yv^\perp\eqdef\left(\frac{1}{\pi'_\xv(x_1)y_1},\dots,\frac{1}{\pi'_\xv(x_n)y_n}\right)
  $$
  and $\pi'_\xv$ is the derivative of $\pi_\xv(x) \eqdef \Pi_{i=1}^n (x-x_i)$.  We also have
  $$
\fsRS{k}^\bot = \fsRS{q-k}
  $$
  \end{proposition}

The Koetter and Vardy decoding algorithm has the following performance. 
\begin{proposition}[\cite{MT17}]\label{Proposition:KV}
	Consider a (generalized) Reed-Solomon code over $\F_q$ of length $n$ and dimension $k$. Let $R = \frac{k}{n}$, a probability function $p : \F_q \rightarrow \mathbb{R}$ and let $C_{KV} = \sum_{e \in \F_q} p^2(e)$. If $R < C_{KV}$ then the Koetter-Vardy algorithm is a decoder (as per Definition\ref{Definition:Decoder}) for $\C$ with error function $p$, that finds with probability $1 - e^{-\Omega((\frac{C_{KV} - R}{R})^2n)}$ a list $L$ of size $\poly(q,n)$ that contains the valid codeword.
\end{proposition}

This means in particular that if $C_{KV}$ is an absolute constant strictly smaller than $R$, then the Koetter-Vardy succeeds wp. $P_{dec} = \frac{1}{\poly(n,d)}$, by choosing a random element of the list $L$. We consider the above decoder but we want it to be deterministic, so we fix beforehand the randomness of the algorithm. Moreover, the Koetter-Vardy decoder is a syndrome decoder.

\subsection{Computational problems related to Reed-Solomon codes}\label{Section:RSProblem}

We now look at specific computational problems related to Reed-Solomon codes that we will study. The first one, which we call $\SRSC$, is a slight special case of the Optimal Polynomial Interpolation introduced in~\cite{JSW+24}. Recall that we consider here Reed-Solomon codes with full support, {\ie} $n = q$.

\begin{problem}[$\SRSC$ problem]\label{Problem:SRSC} $ $ \\
	\textbf{Given: } a uniformly random  $\sv\in \F_q^{q-k}$, an integer $k$ in $\Iint{1}{q}$, a subset 
	$\cS$ of $\fq$, a parity-check matrix $\Hm \in \fq^{(q-k)\times q}$ of a full support Reed-Solomon code of dimension $k$.\\
	\textbf{Goal:} find $\xv = (x_\alpha)_{\alpha \in \fq} \in \F_q^q$ st. $\Hm \trsp{\xv} = \trsp{\sv}$ and $\forall \alpha \in \F_q, \ x_\alpha \in S$.
\end{problem}

This problem can be interpreted as an polynomial interpolation problem. 

\begin{problem}[$\SAINTERPOL$ problem] $ $ \\
	\textbf{Given: } a uniformly random  $\yv=(y_\alpha)_{\alpha \in \fq} \in (\fq)^q$, an integer $k$ in $\Iint{1}{q}$, a fixed subset 
	$\cS$ of $\fq$ \\
	\textbf{Goal:}  find a polynomial $P(X) \in \fq[X]$ of degree $<k$ such that $y_\alpha - P(\alpha) \in \cS$ for 
	any $\alpha \in \fq$.
\end{problem}

An instance of the problem $\SAINTERPOL$ can be transformed in an instance of the 
$\SRSC$ problem by taking $\trsp{\sv} = \Hm \trsp{\yv}$. A solution $\ev$ of the latter problem yields a solution of the first, since $\yv - \ev$ is necessarily
a codeword $\cv$ of $\fsRS{k}$ and therefore of the form $\cv=(P(\alpha))_{\alpha \in \Fq}$. $P$ can be found efficiently by interpolation. The other way round, from
$\sv \in \F_q^{q-k}$ we find by solving a linear system a solution $\yv$ of the linear system
$\Hm \trsp{\yv} = \trsp{\sv}$. A solution $P(X) \in \fq[X]$ of the first problem yields a solution $\ev$ of the second by taking
$\ev_\alpha = y_{\alpha}-P(\alpha)$ for any $\alpha \in \fq$.

A particular case is when $S = \Iint{-u}{u}$. In this case, the problem is actually an $\ISIS_{\infty}$ problem on Reed-Solomon codes.

\begin{problem}[$\RSISIS_\infty$ problem] $ \ $ 
	
	\textbf{Given:} a uniformly random  $\sv\in \F_q^{q-k}$, an integer $k$ in $\Iint{1}{q}$, an integer $u \in \Iint{0}{\frac{q-1}{2}}$, a parity-check matrix $\Hm \in \fq^{(q-k)\times q}$ of a full support Reed-Solomon code of dimension $k$.
	
	\textbf{Goal:} \ find $\xv = (x_\alpha)_{\alpha \in \fq} \in \F_q^q$ st. $\Hm \trsp{\xv} = \trsp{\sv}$ and $\norm{\xv}_{\infty} \le u$.
\end{problem}

We study both the more general problem with a discussion the choice of $S$ ans also put  a special focus on the $\RSISIS_\infty$ problem, where our algorithms will be the most efficient and which will give the best quantum advantage. 

\section{General reduction}
In this section, our goal is to present a general reduction from the syndrome decoding problem of a code $\C$ to a coset sampling problem in the dual code $\C^\bot$. Unless specified otherwise $\C$ is a code defined over a finite $\Fq$ where $q$ is a prime power. 

\subsection{Statement of the theorem}
We refer to the definitions for syndrome decoding and coset sampling presented in Section~\ref{Section:IntroReduction}, as well as the definition of the probability function $u_p^{\Hm,\sv}$. In our setting, we will have a sampler for the coset sampling problem that will have some errors. We define the faithfulness of such an algorithm as follows

\begin{definition}[faithfulness of a $\CSP$ solver]\label{def:faithfulness}
	Consider an algorithm $\aa(\Hm,\sv)$ with $\Hm \in \fq^{(n-k)\times n}$. Let $t_{\sv} : \F_q^n \rightarrow \mathbb{R}_+$ be probability functions defined as
	\begin{align*}
		t_{\sv}(\xv) & \eqdef \Pr[\aa(\Hm,\sv) \textrm{ outputs \xv}]
	\end{align*}
	We say that $\aa$ solves $\CSP(\Hm,p)$ with faithfulness $\gamma$ if
	$\E_{\sv \Unif \F_q^{n-k}}\left[\Delta(t_\sv,u_p^{\Hm,\sv})\right] \le 1 - \gamma.$
\end{definition}

We can now provide the formal statement of our reduction.

\begin{theorem}\label{Theorem:1}
	Let $\C$ be a $q$-ary code specified by the parity-check matrix $\Hm \in \F_q^{(n-k)\times n}$. Let $\Hm^\bot$ be a parity-check  matrix of the dual code $\C^\bot$. Let $F : \F_q^n \rightarrow \mathbb{C}$ with $\norm{F} = 1$. Assume that:
	\begin{enumerate}\setlength\itemsep{-0.2em}
		\item we have access to an algorithm that solves $\SD(\Hm,|F|^2)$ with probability $P_{dec}$, and runs in time $T_{dec}$,
		\item we have access to  a quantum algorithm that outputs the superposition $\sum_{\ev \in \F_q^n} F(\ev) \ket{\ev}$, and runs in time $T_F$.
	\end{enumerate}
	Then, we can construct a quantum algorithm that solves $\CSP(\Hm^{\bot},|\hF|^2)$ with faithfulness $\gamma~=~1~-~\sqrt{1 - P_{dec}}$ which runs in time $O\left(\frac{1}{P_{dec}}\left(T_{dec} + T_F\right)\right)$.
\end{theorem}

Notice that the above works for any parity matrix $\Hm$ (without the need of having random matrices for instance). The only price we pay is that we are in the inhomogeneous setting but as we discussed, such a reduction doesn't hold in the homogeneous setting. Notice also that the second condition, {\ie} being able to efficiently construct $\sum_{\ev \in \F_q^n} F(\ev) \ket{\ev}$ is necessary for all quantum algorithms based on Regev's reduction.

Notice that an algorithm $\aa$ for solving $\SD(\Hm,|F|^2)$ is in particular a syndrome decoder for the code $\C$ associated to the parity matrix $\Hm$, with error $|F|^2$. 

\subsection{A first technical proposition}
The following statement is here to deal with errors in the decoding algorithm. We want to compare the obtained state in the ideal setting (with an error function $F$) after the first step of Regev's reduction with the real setting, with an error function $G$.
\begin{proposition}\label{Proposition:Main}
	Let $\C$ be a linear code described by a parity-check matrix $\Hm \in \F_q^{(n-k)\times n}$. Let $F,G : \F_q^n \rightarrow \mathbb{C}$ such that $\norm{F}_2 = \norm{G}_2 = 1$ where $G$ also satisfies for each $\vv \in \F_q^n$
	\begin {align}\label{Eq:2}
	\norm{ \sum_{\cv \in \C, \ev \in \F_q^n} G(\ev) \car{\cv}{\vv} \ket{\cv + \ev}}^2 = q^k.
\end{align}
Let $\ket{U^F_{\vv}} = \sum_{\cv \in \C, \ev \in \F_q^n} F(\ev) \car{\cv}{\vv} \ket{\cv + \ev}$ and $
Z^F_{\vv} \eqdef \norm{\ket{U^F_{\vv}} }^2$, and $V \eqdef \{\vv \in \F_q^n : Z^F_{\vv} \neq 0\}$. For each $\vv \in \F_q^n$, we define the vectors
\begin{align*}
	\ket{\phi^F_{\vv}}  & \eqdef \frac{1}{\sqrt{Z^F_{\vv}}} \sum_{\cv \in \C, \ev \in \F_q^n} F(\ev) \car{\cv}{\vv} \ket{\cv + \ev} & \textrm{if } \vv \in V \\
	\ket{\phi^F_{\vv}}  & \eqdef \zerov \textrm{ (the null vector) } &  \textrm{otherwise}
\end{align*}
Notice that $\ket{\phi^F_{\vv}}$ is a unit vector for $\vv \in V$. For each $\vv \in \F_q^n$, we also define the unit vectors
$\ket{\phi^G_{\vv}}  \eqdef \frac{1}{\sqrt{q^k}} \sum_{\cv \in \C, \ev \in \F_q^n} G(\ev) \car{\cv}{\vv} \ket{\cv + \ev}.
$
We have
$$ \E_{\vv \Unif \F_q^n}\left[|\braket{\phi_{\vv}^F}{\phi_{\vv}^G}|^2\right]  \ge \left|\sum_{\ev \in \F_q^n} \overline{F(\ev)} G(\ev)\right|^2.$$
\end{proposition}
\begin{proof}
We write for $\vv \in V$.
\begin{align*} \braket{\phi^F_{\vv}}{\phi^G_{\vv}} & = \frac{1}{\sqrt{Z^F_{\vv} q^k}} \sum_{\cv,\cv' \in \C} \sum_{\substack{\ev,\ev' \in \F_q \\ \cv + \ev = \cv' + \ev'}} \car{\cv'-\cv}{\vv}\overline{F(\ev)} G(\ev') \\ 
	& = \frac{1}{\sqrt{Z^F_{\vv} q^k}} \sum_{\cv' \in \C} \sum_{\cv \in \C}\sum_{\ev' \in \F_q} \car{\cv'-\cv}{\vv}\overline{F(\ev' + \cv' - \cv)} G(\ev') \\ 	
	& = \frac{1}{\sqrt{Z^F_{\vv} q^k}} \sum_{\cv' \in \C} \sum_{\cv \in \C}\sum_{\ev' \in \F_q} \car{\cv}{\vv}\overline{F(\ev' + \cv)} G(\ev') \\ 	
	& = \frac{\sqrt{q^k}}{\sqrt{Z^F_{\vv}}} \sum_{\cv \in \C} \sum_{\substack{\ev \in \F_q^n}}  \car{\cv}{\vv}\overline{F(\ev+\cv)}G(\ev)
\end{align*}
We now prove two short lemmata that will then allow us to apply a Cauchy-Schwartz inequality to bound the expected value of $|\braket{\phi^F_{\vv}}{\phi^G_{\vv}}|^2$.
\begin{lemma}
	$ \E_{\vv \Unif \F_q^n}  \left[Z_{\vv}^F\right] = q^k. $
\end{lemma}
\begin{proof}
	Using similar calculations as the ones to compute $\braket{\phi^G_{\vv}}{\phi^F_{\vv}}$, we can write
	\begin{align*}
		Z^F_{\vv} & = \norm{ \sum_{\cv \in \C, \ev \in \F_q^n} F(\ev) \car{\cv}{\vv} \ket{\cv + \ev}}^2 \\
		 & =  \sum_{\cv,\cv' \in \C} \sum_{\substack{\ev,\ev' \in \F_q \\ \cv + \ev = \cv' + \ev'}} \car{\cv'-\cv}{\vv}\overline{F(\ev)} F(\ev') \\ 
		& =  \sum_{\cv' \in \C} \sum_{\cv \in \C}\sum_{\ev' \in \F_q} \car{\cv'-\cv}{\vv}\overline{F(\ev' + \cv' - \cv)} F(\ev') \\ 	
		& =  \sum_{\cv' \in \C} \sum_{\cv \in \C}\sum_{\ev' \in \F_q} \car{\cv}{\vv}\overline{F(\ev' + \cv)} F(\ev') \\ 	
		& = q^k \sum_{\cv \in \C} \sum_{\substack{\ev \in \F_q^n}}  \car{\cv}{\vv}\overline{F(\ev+\cv)}F(\ev).
		\end{align*}
	We then write
	\begin{align*}
		\E_{\vv \Unif \F_q^n}  \left[Z_{\vv}^F\right] & =  \frac{1}{q^n} \sum_{\vv \in \F_q^n} {q^k} \sum_{\cv \in \C} \sum_{\substack{\ev \in \F_q^n}}\car{\cv}{\vv} \overline{F(\ev + \cv)} F(\ev) \\
		& = \frac{q^k}{q^n}  \sum_{\cv \in \C} \left(\sum_{\vv \in \F_q^n}\car{\cv}{\vv}\right) \sum_{\substack{\ev \in \F_q^n}}\overline{F(\ev + \cv)} F(\ev)
	\end{align*}
	Now, notice that $\sum_{\vv \in \F_q^n}\car{\cv}{\vv} = q^n$ for $\cv = 0$ and $\sum_{\vv \in \F_q^n}\car{\cv}{\vv} = 0$ otherwise. This means we can conclude
	\begin{align*}
		\E_{\vv \Unif \F_q^n}  \left[Z_{\vv}^F\right] = q^k \sum_{\ev \in \F_q^n} \overline{F(\ev)} F(\ev) = q^k \norm{F}^2 = q^k.
	\end{align*}
\end{proof}
\begin{lemma}
	Let $\beta_{\vv} \eqdef \sqrt{\frac{1}{q^k}}\braket{U^F_{\vv}}{\phi_{\vv}^G} = \sum_{\cv \in \C, \ev \in \F_q^n} \car{\cv}{\vv} \overline{F(\ev + \cv)} G(\ev)$. We have 
	$$\E_{\vv \Unif \F_q^n}[|\beta_{\vv}|] \ge \left|\sum_{\ev} \overline{F(\ev)} G(\ev)\right|.$$
\end{lemma}	
\begin{proof}
	We write 
	\begin{align*}
		\E_{\vv \Unif \F_q^n}[|\beta_{\vv}|] & \ge \left|\frac{1}{q^{n}} \sum_{\vv \in \F_q^n} \beta_{\vv}\right| = \frac{1}{q^n} \left|\sum_{\cv \in \C} \left(\sum_{\vv \in \F_q^n} \car{\cv}{\vv} \right) \sum_{\ev \in \F_q} \overline{F(\cv + \ev)} G(\ev)\right| \\
		& = \left|\sum_{\ev} \overline{F(\ev)} G(\ev) \right|,
	\end{align*}
	where we used the same argument as in the previous lemma for determining the sum $ \sum_{\vv \in \F_q^n} \car{\cv}{\vv}$.
\end{proof}
We can now conclude. Notice that $|\beta_{\vv}| = \frac{\sqrt{Z^F_{\vv}}}{\sqrt{q^k}} |\braket{\phi_{\vv}^F}{\phi_{\vv}^G}|$ for each $\vv \in \F_q^n$ since $\ket{U^F_{\vv}} = \sqrt{Z^F_{\vv}} \ket{\phi^F_{\vv}}$.
 Using the Cauchy-Schwarz inequality, we have 
$$ \left(\E_{\vv \Unif \F_q^n} \left[|\beta_{\vv}|\right]\right)^2 \le \E_{\vv \Unif \F_q^n} \left[\frac{Z^F_{\vv}}{q^k}\right]\E_{\vv \Unif \F_q^n} \left[|\braket{\phi_{\vv}^F}{\phi_{\vv}^G}|^2\right],$$
which allows us to conclude
$$ \E_{\vv \Unif \F_q^n} \left[|\braket{\phi_{\vv}^F}{\phi_{\vv}^G}|^2\right] \ge q^k \frac{\left(\E_{\vv \Unif \F_q^n} \left[|\beta_{\vv}|\right]\right)^2}{\E_{\vv \Unif \F_q^n} \left[{Z^F_{\vv}}\right]} \ge q^k \frac{\left|\sum_{\ev} \overline{F(\ev)} G(\ev) \right|^2}{q^k}
= \left|\sum_{\ev} \overline{F(\ev)} G(\ev) \right|^2,$$
where we used our two lemmata for the last inequality.
\end{proof}

The constraint on $G$ (Equation~\ref{Eq:2} in the above Proposition) may seem quite stringent. However all decoders we consider here (and most decoders in general) will produce with our algorithm a state $\ket{\phi_{\vv}^G}$ where $G$ satisfies this property, so we will be able to widely use this proposition. This is characterized by the following proposition.

\begin{proposition}\label{Proposition:NormG}
Let $\Hm \in \F_q^{(n-k)\times n}$ be a parity matrix and let $\Dc \subseteq \F_q^n$ such that  the mapping which associates to $\ev \in \Fqn$ its syndrome $\Hm \trsp{\ev}$ is one to one on $\Dc$.
Let $G : \F_q^n \rightarrow \mathbb{C}$ be such that $\Supp(G) \subseteq \Dc$ and $\norm{G} = 1$. Then for each $\vv \in \F_q^n$, we have 
$$ \norm{ \sum_{\cv \in \C, \ev \in \F_q^n} G(\ev) \car{\cv}{\vv} \ket{\cv + \ev}}^2 = q^k.$$
\end{proposition}
\begin{proof}
We write 
\begin{align*}
	\norm{ \sum_{\cv \in \C, \ev \in \F_q^n} G(\ev)  \car{\cv}{\vv} \ket{\cv + \ev}}^2 & =
	\sum_{\cv,\cv' \in \C} \sum_{\substack{\ev,\ev' \in \Dc \\ \cv + \ev = \cv' + \ev'}} \car{\cv-\cv'}{\vv}G(\ev)\overline{G(\ev')}
\end{align*}
Using the property on $\Dc$, we have that for $\cv,\cv' \in \C$ and $\ev,\ev' \in \Dc$, $\left(\cv + \ev = \cv' + \ev' \right) \Rightarrow (\cv,\ev) = (\cv',\ev')$. We can therefore rewrite the above sum as
$$ \norm{ \sum_{\cv \in \C, \ev \in \F_q^n} G(\ev)  \car{\cv}{\vv} \ket{\cv + \ev}}^2  = \sum_{\cv \in \C} \sum_{\ev \in \Dc} G(\ev)\overline{G(\ev)} = q^k \norm{G}^2 = q^k.$$
\end{proof}

This proposition will actually imply that we can use our proposition from any state arising from a syndrome decoder - which can be immediately obtained from an algorithm that solves the $\SD(\Hm,p)$ problem.  

\subsection{Analysis of the general algorithm}\label{Sec:Algorithm}

The goal of this section is now to prove Theorem~\ref{Theorem:1}. We  present the quantum algorithm based on Regev's reduction that performs this reduction. The algorithm uses a syndrome decoder $\aa_{dec}$ (which can be directly obtained from an algorithm $\aa$ that solves $\SD(\Hm,|F|^2)$) and then construct a certain state $\ket{\psi_{\vv}}$. Our dual coset sampler will consist of constructing this state and then measuring it in the computational basis. \\ \\
\cadre{
\begin{center} {\bf Algorithm 1: shifted dual code sampler from a primal decoder} \end{center}
\textbf{Input:} 
\begin{itemize} \setlength\itemsep{-0.2em}
	\item A linear code $\C \in \Fqn$ of dimension $k$ described by a parity-check matrix $\Hm \in \F_q^{n-k \times n}$,
	\item a function $F \in \F_q^n \rightarrow \mathbb{C}$ of norm $1$,
	\item an element $\vv \in \Fqn$.
\end{itemize}
\textbf{Decoder:} a syndrome decoder $\aa_{dec}$ of $\C$ for the error function $|F|^2$ that succeeds with probability $P_{dec}$. 
We denote by $\Dc$ the decoding set of $\aa_{dec}$.\\
\textbf{Goal:} 
Use $\aa_{dec}$ in order to construct a state $\ket{\psi_{\vv}}$ which is not too far from a state proportional to 
$\sum_{\yv \in -\vv+  \C^{\bot}} \hF(\yv) \ket{\yv}$. \\ \\
\textbf{Algorithm:} \\
\begin{tabular}{lll}
	Initial state: & & $\frac{1}{\sqrt{q^k}}\sum_{\cv \in \C} \car{\cv}{\vv}\ket{\cv} \otimes \sum_{\ev \in \C} F(\ev) \ket{\ev}$ \\
	add $\ev$ to $\cv$: & $\mapsto$ & $\frac{1}{\sqrt{q^k}} \sum_{\cv \in \C, \ev \in \F_q^n} F(\ev)  \car{\cv}{\vv} \ket{\cv + \ev} \ket{\ev}$\\
	apply coherently $\aa_{dec}$ & $\mapsto$ & $\ket{\Psi^0_{\vv}}  = \frac{1}{\sqrt{q^k}} \sum_{\cv \in \C, \ev \in \F_q^n} F(\ev)   \car{\cv}{\vv}\ket{\cv + \ev} \ket{\ev - \aa_{dec}(\cv + \ev)}$\\
	measure last register:& & if not $\zero$ start again until getting \\
	& $\mapsto$  & $\ket{\Psi^1_{\vv}}  = \frac{1}{\sqrt{q^k P_{dec}}} \sum_{\cv \in \C, \ev \in \Dc} F(\ev) \car{\cv}{\vv}  \ket{\cv + \ev}$ \\
	apply the QFT: & $\mapsto$ & $\ket{\psi_{\vv}}= \QFT{\ket{\Psi^1_{\vv}}}$
	\end{tabular}}  $ $  \\
	
	\COMMENT{	Construct $\frac{1}{\sqrt{q^k}}\sum_{\cv \in \C} \omega^{\vv \cdot \cv}\ket{\cv} \sum_{\ev \in \C} F(\ev) \ket{- \ev}$ and then add the second register to the first register in order to get the state 
$\frac{1}{\sqrt{q^k}} \sum_{\cv \in \C, \ev \in \F_q^n} F(\ev)  \omega^{\vv \cdot \cv} \ket{\cv + \ev} \ket{- \ev}.$ Apply coherently $\aa_{dec}$ on the two registers to obtain the state 
$$ \ket{\Psi^0_{\vv}}  = \frac{1}{\sqrt{q^k}} \sum_{\cv \in \C, \ev \in \F_q^n} F(\ev)   \omega^{\vv \cdot \cv}\ket{\cv + \ev} \ket{\aa_{dec}(\cv + \ev) - \ev}.$$
Because $\aa_{dec}$ is a syndrome decoder, there is a subset $D \subseteq \F_q^n$ such that $\aa_{dec}(\cv + \ev) = \ev$ for any $\ev \in D$ (for any $\cv$). We measure the last register. If we don't obtain $0^n$, start again. Conditioned on measuring $0^n$, we obtain the state 
$$ \ket{\Psi^1_{\vv}}  = \frac{1}{\sqrt{q^k P_{dec}}} \sum_{\cv \in \C, \ev \in D} F(\ev) \omega^{\vv \cdot \cv}  \ket{\cv + \ev}.$$
We apply the Quantum Fourier Transform on this state and obtain $\ket{\psi_{\vv}}= \QFT{\ket{\Psi^1_{\vv}}}$. }

The following theorem analyzes this algorithm
\begin{theorem}
Let $\C$ be a linear code 
and let $F: \F_q^n \rightarrow \mathbb{C}$ such that $\norm{F}_2  = 1$. Assume we have a syndrome decoder $\aa_{dec}$ that succeeds with probability $P_{dec}$. 
Let $\ket{\Phi_{\vv}} \eqdef \frac{1}{\sqrt{Y^F_{\vv}}} \sum_{\yv \in - \vv +\C^{\bot}} \hF(\yv)\ket{\yv}$, where 
$Y^F_{\vv}$ is a normalizing constant which makes this state to be a unit vector. If $\hF$ is zero on all $-\vv+\C^\perp$, then $\ket{\Phi_{\vv}}$ is defined to be the zero vector.
Then the above algorithm constructs a state $\ket{\psi_{\vv}}$ such that 
$$ \E_{\vv \Unif \F_q^n} \left[|\braket{\Phi_{\vv}}{\psi_{\vv}}|^2\right]  = P_{dec}.$$
\end{theorem}
\begin{proof}
In the algorithm, after applying the decoder $\aa_{dec}$, we have the state 
$$ \ket{\Psi^0_{\vv}} = \frac{1}{\sqrt{q^k}} \sum_{\cv \in \C} \sum_{\ev \in E} F(\ev) \car{\cv}{\vv}\ket{\cv + \ev}\ket{\ev - \aa_{dec}(\cv + \ev)}.$$
Let $\Dc \subseteq \F_q^n$ be the decoding set of $\aa_{dec}$. We can rewrite the above as
$$ \ket{\Psi_0} = \frac{1}{\sqrt{q^k}} \sum_{\cv \in \C} \sum_{\ev \in \Dc} F(\ev) \car{\cv}{\vv} \ket{\cv + \ev}\ket{0^n} + \sum_{\xv \neq \zero} \ket{W_\xv}\ket{\xv}.$$

$\Dc$ has also the property that the mapping $\ev \mapsto \Hm \trsp{\ev}$ is one to one on $\Dc$. Here $\Hm$ denotes any parity-check matrix of $\C$. The success probability of the decoding algorithm is $P_{dec} = \sum_{\ev \in \Dc} |F(\ev)|^2$. 
Let $G$ be the function such that $G(\ev) = \frac{F(\ev)}{\sqrt{P_{dec}}}$ for $\ev \in \Dc$ and $G(\ev) = 0$ otherwise. Using Proposition~\ref{Proposition:NormG}, we have that 
\begin{align}\label{Eq:3}
	\norm{\sum_{\cv \in \C} \sum_{\ev \in \F_q^n} G(\ev) \car{\cv}{\vv} \ket{\cv + \ev}\ket{0^n}}^2 = q^k
\end{align}
and conditioned on measuring $0^n$ in the second register, we obtain the unit vector
$$ \ket{\Psi^1_{\vv}} = \frac{1}{\sqrt{q^k}} \sum_{\cv \in \C} \sum_{\ev \in \F_q^n} G(\ev) \car{\cv}{\vv} \ket{\cv + \ev}.$$
Let $
Z^F_{\vv} \eqdef \norm{ \sum_{\cv \in \C, \ev \in \F_q^n} F(\ev) \car{\cv}{\vv} \ket{\cv + \ev}}^2$, and $V \eqdef \{\vv \in \F_q^n : Z^F_{\vv} \neq 0\}$. For each $\vv \in \F_q^n$, we define the `ideal'' vectors
\begin{align*}
	\ket{\phi^F_{\vv}}  & \eqdef \frac{1}{\sqrt{Z^F_{\vv}}} \sum_{\cv \in \C, \ev \in \F_q^n} F(\ev) \car{\cv}{\vv} \ket{\cv + \ev} & \textrm{if } \vv \in V \\
	\ket{\phi^F_{\vv}}  & \eqdef \zerov \textrm{ (the null vector) } &  \textrm{otherwise}
\end{align*}
Thanks to Equation~\ref{Eq:3}, we can use Proposition~\ref{Proposition:Main} and obtain
$$\E_{\vv \Unif \F_q^n} \left[|\braket{\Psi^1_{\vv}}{\phi^F_{\vv}}|^2\right] \ge \left| \sum_{\ev \in \F_q^n} \overline{F(\ev)} G(\ev)\right|^2 =\left( \frac{1}{\sqrt{P_{dec}}} \sum_{\ev \in D} |F(\ev)|^2\right)^2 = P_{dec}.$$
Also, notice that by using Proposition \ref{proposition:periodic} we have $\QFT{\ket{\phi_{\vv}^F}} = \ket{\Phi_{\vv}}$. Since we also have by definition of $\ket{\psi_{\vv}}$, $\QFT{\ket{\Psi_{\vv}^1}} = \ket{\psi_{\vv}}$, we can conclude  
$$\E_{\vv \Unif \F_q^n} \left[|\braket{\Phi_{\vv}}{\psi_{\vv}}|^2\right] = \E_{\vv \Unif \F_q^n} \left[|\braket{\Psi^1_{\vv}}{\phi^F_{\vv}}|^2\right] \ge P_{dec}.$$
\end{proof}

\begin{corollary}\label{Corollary:Main}
Take the above algorithm for a fixed $\sv \in \F_q^{k}$ a given parity-check matrix $\Hm^\bot$ of 
$\C^\bot$ and any $\vv $ st. $- \Hm^\bot \trsp{\vv} = \trsp{\sv}$, and measure the state $\ket{\psi_{\vv}}$ in the computational basis. Let $t_{\sv}(\yv)$ be the probability of measuring each $\yv$. We have 
$$\E_{\sv \Unif \F_q^{k}}\left[\Delta(t_\sv,u^{\Hm^{\bot},\sv}_{|F|^2}) \le \sqrt{1 - P_{dec}}\right].$$
\end{corollary}
\begin{proof}
Using the above notation, we have that $\QFT{\ket{\phi_{\vv}^F}} = \ket{\Phi_{\vv}}$ is a unit vector proportional to $\sum_{\yv \in -\vv + \C^{\bot}} \hF(\yv)\ket{\yv}$ which means that if  this state is measured in the computational basis, we get $\yv$ with probability $u^{\Hm^{\bot},\sv}_{|F|^2}(\yv)$. Now the algorithm produces a state $\ket{\psi_{\vv}}$, and measuring it in the computational basis gives  $\yv$ with probability $t_\sv$. From there, we have 
$$ \Delta(t_\sv,u^{\Hm^{\bot},\sv}_{|\hF|^2}) \le \Delta(\ket{\Phi_{\vv}},\ket{\psi_{\vv}}),$$
where we use the fact that the trace distance of two quantum pure states $\sum_{x} a(x) \ket{x}$ and $\sum_{x} b(x) \ket{x}$ is at least as large as the trace distance between $|a|^2$ and $|b|^2$ (see~\cite{NC00} for instance).
This allows us to conclude 
$$\E_{\sv \Unif \F_q^{k}} \left[\Delta(t_\sv,u^{\Hm^{\bot},\sv}_{|\hF|^2})\right] \le \E_{\sv \Unif \F_q^{k}} \left[\sqrt{1 - |\braket{\Phi_{\vv}}{\psi_{\vv}}|^2}\right] \le
\sqrt{1 - \E_{\sv \Unif \F_q^{k}}\left[|\braket{\Phi_{\vv}}{\psi_{\vv}}|^2\right]} = \sqrt{1 - P_{dec}}.$$
where we use Jensen's inequality and the concavity of $x \rightarrow \sqrt{1 - x^2}$ on $[0,1]$. This proves our corollary.
\end{proof}

\paragraph{Running time of the algorithm.}
Let $T_F$ be the time required to compute $\sum_{\ev} F(\ev)\ket{\ev}$. Let $T_{\aa_{dec}}$ be the running time of the decoder. The above algorithm runs in average time
$O\left(\frac{1}{P_{dec}}(T_F+ T_{\aa_{dec}})\right)$. The corollary and the running time give immediately Theorem~\ref{Theorem:1}. \\

Notice that the above can be improved to $O(\frac{1}{\sqrt{P_{dec}}} (T_F + T_A))$ if needed using quantum amplitude amplification instead of measuring just the final register and conditioning on obtaining $0^n$. We work in regimes where there this has little impact but could be an interesting remark in other settings.

\subsection{Solving the $\SAINTERPOL$ problem by using the KV decoder}

We will now instantiate the above algorithm with Reed-Solomon codes using the Koetter-Vardy decoder. Our problem of interest will be the $\SRSC$ problem defined in Section~\ref{Section:RSProblem}, which is equivalent to $\SAINTERPOL$. Recall that since we consider full support Reed-Solomon codes, we have $n = q$. We consider the following algorithm \\ \\
\cadre{
\begin{center}{\bf Algorithm 2 solving $\SRSC$} \end{center}
\textbf{Input:} A parity-check matrix $\Hm' \in \F_q^{(q-k)\times q}$ of $\C' = \fsRS{k}$, a subset 
$\cS$ of $\fq$, a random syndrome $\sv \in \F_q^{q-k}$. \\
\textbf{Parameter: } a function $f : \F_q \rightarrow \mathbb{C}$ such that $\Supp(\hf) \subseteq \cS$ and $\norm{f} = 1$. \\
\textbf{Algorithm:}
\begin{enumerate}
	\item Let $\Hm \in \F_q^{k,q}$ be a parity matrix of $\C = (\C')^\bot$. Run Algorithm 1 on $\Hm$ with $n = q, k = q-k'$, with $F = f^{\otimes n}$, and  using the Koetter-Vardy decoder $\aa_{KV}$.
	\item Let $\yv$ be the output of Algorithm $1$ with these parameters. If $\yv$ is a correct solution to our $\SRSC$ instance, output $\yv$, otherwise repeat.
\end{enumerate}
} $ \ $ \\

\begin{theorem}\label{Theorem:ISISf}
The above quantum algorithm solves in polynomial time  the  $\SRSC$ problem as long as $\frac{k'}{q} > 1 -  \sum_{\alpha \in \F_q} |f(\alpha)|^4$.
\end{theorem}
\begin{proof}
	Using notations from Section~\ref{Sec:Algorithm}, notice that if we sample according $u^{H',\sv}_{|\hF|^2}$ then we always output a valid solution $\yv$. From Corollary~\ref{Corollary:Main}, we output a valid solution $\yv$ wp. at least $1 - \sqrt{1 - P_{dec}}$ on average on $\sv$. $P_{dec}$ here is the success probability of the $KV$ decoder. From Proposition~\ref{Proposition:KV}, we know that $P_{dec} \ge \frac{1}{\poly(q)}$ as long as $\frac{k}{q} < \sum_{\alpha \in \F_q} |f(\alpha)|^4$ which implies $\frac{k'}{q} > 1 -\sum_{\alpha \in \F_q} |f(\alpha)|^4$. When this happens, we will repeat in step $2$ at most $\poly(q)$ times and our whole algorithm will run in quantum polynomial time.  
\end{proof}

This raises the following question. What function $f$ should we use such that $\Supp(\hf) \subseteq \cS$. After giving a general formula for $\sum_{\alpha \in \F_q} |f(\alpha)|^4$ as a function of $\hf$, we will treat the important case $\RSISIS_\infty$ in detail. Then, we will consider more general sets $S$.

\subsection{A general formula for $\sum_{\alpha \in \F_q}|f(\alpha)|^4$.}

Parseval's identity gives
$ \sum_{\alpha \in \F_q} |f (\alpha)|^2  =  \sum_{\alpha \in \F_q} |\hf (\alpha)|^2$, but we are interested in fourth powers of $|f|$ so we cannot use this directly. But still a suitable use of Parseval's identity yields
\begin{proposition}
\label{prop:p2}
Let $p \eqdef |f|^2$. We have
$$
\sum_{x \in \F_q} p^2(x) = \sum_{x \in \F_q} |f(x)|^4 = \frac{1}{q} \sum_{x \in \F_q} |\hf \star \hf(x)|^2.
$$
\end{proposition}
\begin{proof}
We have
\begin{eqnarray*}
	\sum_{x \in \F_q} p^2(x) &= &\sum_{x \in \F_q} |f(x)|^4 \\
	& = & \sum_{x \in \F_q} |f^2(x)|^2 \\
	& = & \sum_{x \in \Fq} |\widehat{f^2}(x)|^2 \;\;\text{(by Parseval applied to $f^2$)}\\
	& = & \frac{1}{q}\sum_{x \in \Fq} \left| \hf \star \hf (x) \right|^2 \;\;\text{(since $\widehat{f \cdot f}= \frac{1}{\sqrt{q}} \hf \star \hf$)}
\end{eqnarray*}
\end{proof}

We will in particular use this formula when $\hf$ is the indicator function of a set $\cS$.
\begin{corollary}
\label{cor:indicator}
Let $f$ be such that $\hf = \one_\cS$ and $p \eqdef |f|^2$.
We have
$$
\sum_{x \in \Fq} p^2(x) =  \frac{1}{q} \sum_{x \in \Fq} N_{\cS}^2(x),
$$
where $N_\cS(x) = |\{(a,b) \in \cS \times \cS: a+b=x\}$.
\end{corollary}
\begin{proof}
We obtain by using Proposition \ref{prop:p2} 
\begin{eqnarray*}
	\sum_{x \in \Fq} p^2(x)  =  \frac{1}{q} \sum_{x \in \F_q} |\hf \star \hf(x)|^2 
	= \frac{1}{q} \sum_{x \in \F_q} |\one_\cS \star \one_\cS(x)|^2 
	=  \frac{1}{q} \sum_{x \in \Fq} N^2_{\cS}(x),
\end{eqnarray*}
where the last equality comes from the fact that $\one_\cS \star \one_\cS(x) = \sum_{a \in \Fq} \one_\cS(a)\one_\cS(x-a)=N_\cS(x)$.
\end{proof}
\section{Computing the efficiency of our algorithm for $\RSISIS_\infty$}
We are in the setting where we have the constraint set $S = \Iint{-u}{u}$. We know from Theorem~\ref{Theorem:ISISf} that by taking a function $f : \F_q \rightarrow \mathbb{C}$ st. $\Supp(\hf) \subseteq \Iint{-u}{u}$ and $\norm{f} = 1$, our quantum algorithm solves $\RSISIS_\infty(q,'k,u)$ as long as $\frac{k'}{q} > 1 - \sum_{\alpha \in \F_q} |f(\alpha)|^4$. Moreover, we have from the previous section that 
$$ \sum_{\alpha \in \F_q} |f(\alpha)|^4 = \frac{1}{q}\sum_{x \in \Fq} \left| \hf \star \hf (x) \right|^2.
$$
Our goal is hence to find a function $f$ st. $\hf$ satisfies the support constraint with the above quantity is as large as possible.
\subsection{The uniform function}
A natural choice of function is the function $f$ such that $\hf = \frac{1}{\sqrt{2u + 1}} \one_{\Iint{-u}{u}}$.
We now distinguish two cases, the case where $u \le \frac{q-1}{4}$ and the case $u > \frac{q-1}{4}$.

\begin{proposition}\label{Proposition:6}If $u \le \frac{q-1}{4}$ then
	$$\sum_{x \in \F_q} |f(\alpha)|^4 = \frac{2}{3} \frac{2u+1}{q} + \frac{1}{3q(2u+1)} \ge \frac{2}{3}\rho,$$
	where $\rho = \frac{|S|}{q} = \frac{2u+1}{q}$.
\end{proposition}
\begin{proof}
	By using Corollary \ref{cor:indicator} we know that
	$$ 
	\sum_x |f(\alpha)|^4= \frac{1}{q(2u+1)^2} \sum_{x \in \F_q} |\one_\cS \star \one_\cS(x)|^2  =\frac{1}{q(2u+1)^2} \sum_{x \in \Fq} N^2_{\Iint{-u}{u}}(x).
	$$
	One can check that
	$$
	N_{\Iint{-u}{u}}(x) = \left\{
	\begin{array}{cl}
		2u + 1 - |x|& \textrm{ if } x \in \{-2u,\dots,2u\} \\
		0 & \textrm{ otherwise }
	\end{array} \right.	
	$$
	From this we deduce
	\begin{align*}
		\sum_x |f(\alpha)|^4 & = \frac{1}{q(2u + 1)^2} \sum_{x=-2u}^{2u} (2u + 1 - |x|)^2 \\
		& = \frac{1}{q(2u + 1)^2} \left((2u + 1)^2 + 2 \sum_{x=1}^{2u} (2u + 1 - x)^2\right) \\
		& = \frac{1}{q(2u + 1)^2} \left((2u + 1)^2 + 2 \sum_{x=1}^{2u} x^2\right) 
	\end{align*}
	We now use the equality $\sum_{x = 1}^{2u} x^2 = \frac{1}{3}u(2u+1)(4u+1)$ to obtain
	$$
	\sum_x |f(\alpha)|^4= \frac{1}{q(2u + 1)} \left((2u + 1) + \frac{2u(4u+1)}{3}\right) = 
	\frac{2}{3}\frac{4u^2 + 4u + \frac{3}{2}}{q(2u + 1)}.$$
	Finally, we use the equality $4u^2 + 4u + \frac{3}{2} = (2u+1)^2 + \frac{1}{2}$ to obtain the simplified expression 
	$$
	\sum_x |f(\alpha)|^4 = \frac{2}{3}\frac{2u  + 1}{q} + \frac{1}{3q(2u+1)} \ge \frac{2}{3} \rho.$$ 
\end{proof}

We now proof the case of large $u$.
\begin{proposition}\label{Proposition:7} If $u > \frac{q-1}{4}$ then
	$$\sum_{x \in \F_q}|f(\alpha)|^4  \ge \frac{10}{3}\rho - 4 + \frac{2}{\rho} - \frac{1}{3\rho^2},$$
	where $\rho = \frac{2u+1}{q}$. 
\end{proposition}
\begin{proof}
	In this case, $2u > \frac{q-1}{2}$ so the calculation of
	$N_{\Iint{-u}{u}}(x)$ differs.
	Let $\ell \eqdef q - (2u + 1)$. We distinguish $2$ cases:
	\begin{enumerate}
		\item For $x \in \Iint{-\ell}{\ell}$, we have $N_{\Iint{-u}{u}}(x) = (2u + 1) - |x|$.
		\item For $x \notin \Iint{-\ell}{\ell}$, we have $N_{\Iint{-u}{u}}(x) = (2u + 1) - \ell$.
	\end{enumerate}
	
\noindent We have here $\sum_x |f(\alpha)|^4  = \frac{1}{q(2u + 1)^2} \sum_{x \in \Fq} N^2_{\Iint{-u}{u}}(x)$ which gives
	\begin{align}\label{Eq:BigEq}
		\sum_x |f(\alpha)|^4
		& = \frac{1}{q(2u + 1)^2} \left((2u + 1)^2  + (q - (2\ell + 1))(2u + 1 - \ell)^2 + 2 \sum_{x = 1}^{\ell} (2u + 1 - x)^2\right) 
	\end{align}
	We write $\rho = \frac{2u + 1}{q}$. 
	The calculations are a little bit cumbersome, so we separate them in different lemmata. We start from the rightmost term, 
	\begin{lemma}
		$2 \sum_{x = 1}^{\ell} (2u + 1 - x)^2 = \rho^3 (-\frac{14q^3}{3}) + \rho^2 (8q^3 + 3q^2) + \rho(-4q^3 - 4q^2 - \frac{q}{3}) + \frac{2q^3}{3} + q^2 + \frac{q}{3}.$
	\end{lemma}
	\begin{proof}
		\begin{align*} 
			2 \sum_{x = 1}^{\ell} (2u + 1 - x)^2 = \underbrace{2 \sum_{x = 1}^l x^2}_{=A} - \underbrace{4(2u+1)\sum_{x = 1}^l x}_{=B}  + \underbrace{2l(2u+1)^2}_{=C}\end{align*}
		We now write each of these three terms as a polynomial in $\rho$. We first write (using $l = q(1-\rho)$)
		\begin{align*}
			A & = \frac{1}{3} l(l+1)(2l+1) = \frac{2}{3}l^3 + l^2 + \frac{l}{3} \\
			& = \frac{2q^3}{3}(1-\rho)^3 + q^2(1-\rho)^2 + \frac{q(1-\rho)}{3} \\
			& = \rho^3(-\frac{2q^3}{3}) + \rho^2(2q^3 + q^2) + \rho (-2q^3 - 2q^2 - \frac{q}{3}) + \frac{2q^3}{3} + q^2 + \frac{q}{3}
		\end{align*}
		We now write, 
		\begin{align*}
			B & = 4(q\rho)\frac{l(l+1)}{2} = 2q\rho l^2 + 2q\rho l \\
			& = 2q^3\rho(1-\rho)^2 + 2q^2 \rho(1-\rho) \\
			& = \rho^3 (2q^3) + \rho^2(-4q^3 - 2q^2) + \rho(2q^3 + 2q^2)
		\end{align*}
		Finally, we write 
		\begin{align*}
			C & = 2q^3\rho^2(1-\rho) = \rho^3(-2q^3) + \rho^2(2q^3)
		\end{align*}
		In order to conclude, we write 
		\begin{align*} 
			2 \sum_{x = 1}^{\ell} (2u + 1 - x)^2 & = A - B + C \\
			& = \rho^3 (-\frac{14q^3}{3}) + \rho^2 (8q^3 + 3q^2) + \rho(-4q^3 - 4q^2 - \frac{q}{3}) + \frac{2q^3}{3} + q^2 + \frac{q}{3}
		\end{align*}
	\end{proof}
	We now go and characterize the rest of the term inside the parenthesis in Equation~\ref{Eq:BigEq}.
	\begin{lemma}
		$$(2u + 1)^2  + (q - (2\ell + 1))(2u + 1 - \ell)^2 = \rho^3 (8q^3) + \rho^2(-12q^3 - 3q^2) + \rho(6q^3 + 4q^2) - q^3 - q^2. $$
	\end{lemma}
	\begin{proof}
		Let $X$ be the above term. We write 
		\begin{align*}
			X & = q^2\rho^2 + (2q\rho - (q+1))(2q\rho - q)^2 \\
			& =  q^2\rho^2 + q^2(2q\rho - (q+1))(4\rho^2 - 4\rho + 1) \\
			& = \rho^3 (8q^3) + \rho^2(q^2 - 8q^3 -4q^2(q+1)) + \rho(2q^3 + 4q^2(q+1)) - q^2(q+1) \\
			& = \rho^3 (8q^3) + \rho^2(-12q^3 - 3q^2) + \rho(6q^3 + 4q^2) - q^3 - q^2
		\end{align*}
	\end{proof}
	We can now put everything together, we write 
	\begin{align*}
		\sum_x |f(\alpha)|^4 & = \frac{1}{q(2u + 1)^2} \left((2u + 1)^2  + (q - (2\ell + 1))(2u + 1 - \ell)^2 + 2 \sum_{x = 1}^{\ell} (2u + 1 - x)^2\right) \\
		& = \frac{1}{q^3 \rho^2} \left(\rho^3(\frac{10q^3}{3}) + \rho^2(-4q^3) + \rho(2q^3 - \frac{q}{3}) - \frac{q^3}{3} + \frac{q}{3} \right) \\
		& = \frac{10}{3}\rho - 4 + \frac{2}{\rho} - \frac{1}{3\rho^2} + \frac{q}{3}(1-\rho) \\
		& \ge \frac{10}{3}\rho - 4 + \frac{2}{\rho} - \frac{1}{3\rho^2}.
	\end{align*}
\end{proof}

Combining Propositions~\ref{Proposition:6},\ref{Proposition:7} and Theorem~\ref{Theorem:ISISf}, we immediately obtain the following theorem

\begin{theorem}\label{Theorem:RSISIS}
	There exists a quantum algorithm that solves $\RSISIS_\infty$ with parameters $q,k',u$ as long as $k' \ge V(\rho)n$ where $\rho = \frac{2u+1}{q}$ and 
	\begin{align*}
		V(\rho) = \left\{ 
		\begin{array}{ll}
			1 - \frac{2\rho}{3} & \textrm{for } \rho < \frac{1}{2} \\
			5 - \frac{10\rho}{3} - \frac{2}{\rho} + \frac{1}{3\rho^2} & \textrm{for } \rho > \frac{1}{2}
		\end{array} \right.
	\end{align*}
\end{theorem}

\COMMENT{\begin{figure}
		\begin{center}
			\includegraphics[width = 10cm]{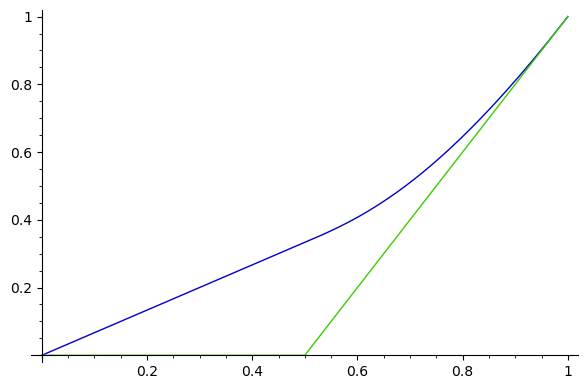} \end{center}
		\caption{Admissible rate $R$ as a function of $\frac{2u+1}{q} = \frac{|S|}{q}$. The blue line corresponds to using our algorithm with uniform $\hf$ and the green corresponds to using Berlekamp-Welch.
			\label{fig:admissible}}
	\end{figure}
}

\subsection{Non-uniform functions and the quest for the optimal function}
Recall, that our goal is to find functions $f : \F_q \rightarrow \mathbb{C}$ s.t. $\Supp(f) \subseteq \Iint{-u}{u}$ with $\norm{f}^2 (= \norm{\hf}^2) = 1$ that maximizes the quantity $\sum_{\alpha \in \F_q} |f(\alpha)|^4 = \frac{1}{q}\sum_{\alpha \in \F_q} |\hf \star \hf|^2(\alpha)$. While this general optimization problem on complex functions seems out of reach, we provide a few insights on this problem. We start from a simple lemma:
\begin{lemma}
	Let $f : \F_q \rightarrow \mathbb{C}$ s.t. $\hf \in \Iint{-u}{u}$ and let $g$ such that $\hg = |\hf|$. Then $\sum_{x} |\hg \star \hg|^2(x) \ge \sum_{x} |\hf \star \hf|^2(x)$.
\end{lemma}
\begin{proof}
	For each $x$ we write 
	$$ |\hf \star \hf|(x) = |\sum_{y \in \F_q} \hf(y)\hf(x-y)| \le \sum_{y \in \F_q} |\hf(y)\hf(x-y)| = \sum_{y \in \F_q} \hg(y)\hg(x-y) = |\hg \star \hg|(x).$$
	From there, $\sum_{x} |\hg \star \hg|^2(x) \ge \sum_{x} |\hf \star \hf|^2(x)$ easily follows.
\end{proof}
This means we only have to look for functions $f$ such that $\hf$ is real and non-negative. In the previous section, we considered the function $\hf(x) = \frac{1}{\sqrt{2u+1}}$ for $x \in \Iint{-u}{u}$ and $\hf(x) = 0$ otherwise. Our goal here is to show that this is, in general, not the optimal function. 

\subsubsection{$u = 1$ and $q \ge 5$.} We first look at the case $u = 1$ and $q=5$. This means we look for functions with $\Supp(\hf) \in \{-1,0,1\}$. In this case, we have 
\begin{align*}
	(\hf \star \hf)(0) & = \hf^2(0) + 2\hf(1)\hf(-1) \\
	(\hf \star \hf)(1) & = 2 \hf(0)\hf(1) \\
	(\hf \star \hf)(2) & = \hf^2(1)\\
	(\hf \star \hf)(-1) & = 2 \hf(0)\hf(-1) \\
	(\hf \star \hf)(-2) & = \hf^2(-1)
\end{align*}

When we consider the balanced function $\hf(0) = \hf(-1) = \hf(1) = \frac{1}{\sqrt{3}}$, we have 
\begin{align*}
	(\hf \star \hf)(0)  = 1 \quad ;\quad 
	(\hf \star \hf)(1) =  (\hf \star \hf)(-1)  = \frac{2}{3} \quad ; \quad
	(\hf \star \hf)(2) =  (\hf \star \hf)(-2)  = \frac{1}{3} 
\end{align*} 
which gives 
$$ \sum_{x \in \F_q} |f \star f|^2(x) = \frac{19}{9}.$$

Now, if we optimize, one can check that the optimal value is obtained by taking $\hf(0) = \sqrt{\frac{3}{7}}$ and $\hf(1) = \hf(-1) = \sqrt{\frac{2}{7}}$. With these values, we obtain
\begin{align*}
	(\hf \star \hf)(0)  = 1 \quad ;\quad 
	(\hf \star \hf)(1) =  (\hf \star \hf)(-1)  = \frac{2\sqrt{6}}{7} \quad ; \quad
	(\hf \star \hf)(2) =  (\hf \star \hf)(-2)  = \frac{2}{7} 
\end{align*} 
which gives

$$ \sum_{x \in \F_q} |\hf \star \hf|^2(x) = 1 + \frac{48}{49} + \frac{8}{49} = \frac{15}{7} > \frac{19}{9} .$$

\subsubsection{General numerical results}
What the intuition shows, and what is confirmed by the previous results is that we want a function $\hf$ which is more concentrated in the center {\ie} around $0$. This comes from the fact that the values $\hf(x)$ for $x$ around $0$ appear more in the decomposition $\hf \star \hf(y)$.

Here, we exhibit one of these functions. We consider the case $q = 1009$ and the function 
$ \hf(x) =N (1 + u - |x|)^{\frac{1}{6}}$ for $x \in \Iint{-u}{u}$ and $\hf(x) = 0$ otherwise, where $N$ is a normalization factor so that $\norm{\hf} = 1$. This is a function which is somewhat concentrated towards the center. With this function, we plot the minimum admissible rate $\frac{k'}{n}$ our algorithm works for $\RSISIS_\infty$ as a function of $\rho = \frac{2u+1}{q}$. We compare this with the case of the uniform function 

\begin{figure}[!ht]
	\begin{center}
		\includegraphics[width = 6.66cm]{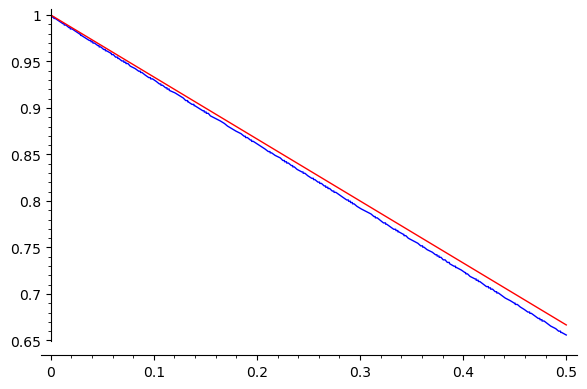} \end{center}
	\caption{Admissible rate $R' = \frac{k'}{n}$ as a function of $\rho = \frac{2u+1}{q}$. The blue line corresponds to the function described above and the red line corresponds to the uniform distribution. These are numerical results with $q = 1009$.}
	\label{Figure:Num}
\end{figure}

This plot shows that the proposed function has slightly better performances than the uniform function. We have tested different decreasing functions and several give slight improvements as the one we presented. 

\section{Approximate interpolation}\label{sec:generalS}
We now consider the more general $\SAINTERPOL$ problem. Again, we will be guided here by Proposition~\ref{prop:p2}. The structure of $\cS$ will play an important role. As a warm-up, we consider the case where $\cS$ is without collisions in the $2$-sum.  

\subsection{The case when $\cS$ is without collisions in the $2$-sums}
A set $\cS$ without collisions in the $2$-sums is such that all sums $a+b$ with $a \leq b$ and $a$ and $b$ in 
$\cS$ are different. This problem can be solved quantumly by our shifted dual code sampler based on the KV decoder of Reed-Solomon codes. We will look for a function $f : \fq \rightarrow \mathbb{C}$ 
of norm $1$ which is such that $\Supp(\hf)\in \cS$ and $\hf$ is non-negative. 

What we will show is that our algorithm works much better in the case of $\ISIS_{\infty}$ than in this setting, showing the importance of the structure of $\cS$.

\begin{lemma}
	Let $\cS$ be a set without collisions in the $2$-sums. Let $g$ be a nonnegative function of unit norm and supported on $\cS$. We have
	$$
	\frac{1}{q} \sum_{x \in \fq} \left| g \star g(x)\right|^2 \leq \frac{2|\cS| - 1}{q|\cS|},
	$$
	and this value is attained for $g = \frac{1}{\sqrt{|\cS|}} \one_{\cS}$.
\end{lemma}
\begin{proof}
	Because of the property on $\cS$ we have that 
	$$ g \star g (x) = \left\{
	\begin{array}{cl}
		g^2(\frac{x}{2}) & \textrm{ if } \frac{x}{2} \in \cS \\
		2g(a)g(b) & \textrm{ if } x = a+b \textrm{ with } a,b > a \in \cS  \\
		0 & \textrm{ otherwise }
	\end{array}
	\right.$$
	From there, we obtain
	\begin{align*}
		\sum_{x \in \fq} \left| g \star g(x)\right|^2 & = \sum_{a \in \cS} g(a)^4 + 4 \sum_{a,b \in \cS, a <b} g(a)^2g(b)^2.
	\end{align*}
	Moreover, 
	\begin{eqnarray}
		\sum_{a \in \cS} g(a)^4 + 4 \sum_{a,b \in \cS, a <b} g(a)^2g(b)^2 & = & 2 \left( \sum_{a \in \cS} g(a)^2\right)^2
		-\sum_{a \in \cS} g(a)^4 \nonumber \\
		& \leq & 2 - \frac{\left(\sum_{a\in \cS} g(a)^2 \right)^2}{|\cS|} \label{eq:CS}\\
		& = & 2 - \frac{1}{|\cS|},
	\end{eqnarray}
	where \eqref{eq:CS} follows from the Cauchy-Schwartz inequality $|\braket{\yv}{\one_\cS}|^2 \leq \norm{\yv}^2
	\norm{\one_\cS}^2$ applied to $\yv=(y_a)_{a \in \fq}$ where $\yv_a = g(a)^2$ if $a \in \cS$ and $0$ otherwise. From there, we obtain 
	$$ \frac{1}{q} \sum_{x \in \fq} \left| g \star g(x)\right|^2 \le \frac{1}{q}\left(2 - \frac{1}{|\cS|}\right) = \frac{2|\cS| - 1}{q|\cS|}.$$
	This gives the upper-bound and the fact that it can be attained can be either verified by taking $g$ to be constant on $\cS$ or by noting that the Cauchy-Schwartz inequality is actually an equality precisely in this case.
\end{proof}

The above show that in this setting, the optimal function $\hf$ is the uniform function on $\cS$, in which case $\frac{1}{q} \sum_{x \in \fq} \left| g \star g(x)\right|^2 = o(1)$ as $q,|S| \rightarrow \infty$. A natural question is to ask whether this also holds say for random $\cS$. As we show in the next section, our algorithm actually has a pretty good performance in the case of a random $\cS$.

\subsection{The case of a random $\cS$}
The previous example showed the worst case for our algorithm. Here, we present the case of a random $\cS$ which is probably more natural. Again, we only analyze here the function $f$ st. $\hf = \frac{1}{\sqrt{|\cS|}} \one_{\cS}$. We begin with a lemma
\begin{lemma}\label{lem:random_S}
	When $\cS$ is chosen uniformly at random, we have as $q$ and $s \eqdef |\cS|$ tend to infinity
	\begin{eqnarray*}
		\esp_\cS \left[ \sum_{x \in \fq} \left| \one_\cS \star \one_\cS(x)\right|^2\right]& = & \frac{s^4}{q} + \OO{\frac{s^3}{q}} \\
		\var_\cS \left( \sum_{x \in \fq} \left| \one_\cS \star \one_\cS(x) \right|^2\right)&=& 
	\OO{\frac{s^7}{q^2}}
	\end{eqnarray*}
\end{lemma}
\begin{proof}
	We have
	\begin{eqnarray}
		\esp_\cS \left[ \sum_{x \in \fq} \left| \one_\cS \star \one_\cS(x)\right|^2\right]
		& = & \esp_\cS \left[ \sum_{x \in \fq} \left| 
		\sum_{\substack{a,b \in \fq: \\a+b=x}} \one_{\cS}(a)\one_{\cS}(b)\right|^2\right] \nonumber\\
		&=& \sum_{x \in \fq} \sum_{\substack{a,b \in \fq: \\a+b=x}}\sum_{\substack{c,d \in \fq: \\c+d=x}}
		\esp_\cS\left\{\one_{\cS}(a)\one_{\cS}(b)\one_{\cS}(c)\one_{\cS}(d)\right\} \label{eq:sum_esp}
	\end{eqnarray}
	In this sum, for a given $x$ and the $4$-tuples $(a,b,c,d)$ such that 
	$a+b=c+d=x$, we distinguish between $4$ cases:\\
	{\bf Case 1:} $a,b,c,d$ are distinct. There are 
	$(q-1)(q-3)$ $4$-tuples of this kind. Indeed, one can choose any $a \in \F_q$ different from $\frac{x}{2}$ (otherwise we have $b = x - a = a$). We then choose $c \in \F_q \backslash \{a,b,\frac{x}{2}\}$, which gives $(q-1)(q-3)$ tuples in total. We have in this 
	case $\esp_\cS\left[\one_{\cS}(a)\one_{\cS}(b)\one_{\cS}(c)\one_{\cS}(d)\right]
	= \frac{\binom{s}{4}}{\binom{q}{4}}=\frac{s^4}{q^4} + \OO{\frac{s^3}{q^4}}$.\\
	{\bf Case 2:} there are $3$ distinct elements among $\{a,b,c,d\}$ and either $a=b$ or $c=d$.
	There are $2(q-1)$ tuples of this kind.
	We have in this 
	case $\esp_\cS\left[\one_{\cS}(a)\one_{\cS}(b)\one_{\cS}(c)\one_{\cS}(d)\right]
	= \frac{\binom{s}{3}}{\binom{q}{3}}=\frac{s^3}{q^3} + \OO{\frac{s^2}{q^3}}$.\\
	{\bf Case 3:} there are $2$ distinct elements among $\{a,b,c,d\}$.
	There are $2(q-1)$ tuples of this kind.
	We have in this 
	case $\esp_\cS\left[\one_{\cS}(a)\one_{\cS}(b)\one_{\cS}(c)\one_{\cS}(d)\right]
	= \frac{\binom{s}{2}}{\binom{q}{2}}=\frac{s^2}{q^2}+ \OO{\frac{s}{q^2}}$.\\
	{\bf Case 4:} $a=b=c=d$.
	There is just one $4$-tuple of this kind.
	We have in this 
	case $\esp_\cS\left[\one_{\cS}(a)\one_{\cS}(b)\one_{\cS}(c)\one_{\cS}(d)\right]
	=\frac{s}{q}$.\\
	By using this classification and computations in \eqref{eq:sum_esp}, we obtain 
	$$
	\esp_\cS \left[ \sum_{x \in \fq} \left| \one_\cS \star \one_\cS(x)\right|^2\right] =  \frac{s^4}{q}+ \OO{\frac{s^3}{q}}.
	$$

Concerning now the variance, we may observe that the computation we have done generalizes easily and that we have
	\begin{eqnarray}
		\esp_\cS \left[ \left(\sum_{x \in \fq} \left| \one_\cS \star \one_\cS(x)\right|^2\right)^2 \right] &= & 
		\sum_{x,x' \in \fq} \sum_{\substack{a,b,c,d \in \fq: \\a+b=x \\c+d=x}}
		\sum_{\substack{a',b',c',d' \in \fq: \\a'+b'=x' \\c'+d'=x'}}
		\esp_\cS\left[\one_{\cS}(a)\one_{\cS}(b)\one_{\cS}(c)\one_{\cS}(d)\one_{\cS}(a')\one_{\cS}(b')\one_{\cS}(c')\one_{\cS}(d')\right] 
		\label{eq:nightmare}
	\end{eqnarray}
	The dominant term in this sum comes from all terms $a,b,c,d,a',b',c',d'$ which are different. The term $\esp_\cS\left[\one_{\cS}(a)\one_{\cS}(b)\one_{\cS}(c)\one_{\cS}(d)\one_{\cS}(a')\one_{\cS}(b')\one_{\cS}(c')\one_{\cS}(d')\right]$ is equal to $\frac{\binom{s}{8}}{\binom{q}{8}}= \frac{s^8}{q^8} + \OO{\frac{s^7}{q^8}}$ in this case. Similarly for a 
	given pair $(x,x')$ there are $q^4 + \OO{q^3}$ $8$-tuples of this kind in the sum and $q^2+\OO{q}$ pairs $(x,x')$. This 
	contributes to a total of $\frac{s^8}{q^2} + \OO{\frac{s^7}{q^2}}$ in the sum. All the other terms can be shown similarly to what has been done for the previous 
	computation to contribute at most $\OO{\frac{s^7}{q^2}}$ to the sum.
	
	Concerning the variance we have
	\begin{eqnarray*}
		\var_\cS \left( \sum_{x \in \fq} \left| \one_\cS \star \one_\cS(x) \right|^2\right)&=& 
		\esp_\cS \left[ \left(\sum_{x \in \fq} \left| \one_\cS \star \one_\cS(x)\right|^2\right)^2 \right]
		- \left( \esp_\cS \left[ \sum_{x \in \fq} \left| \one_\cS \star \one_\cS(x)\right|^2\right]\right)^2
		\\
		& = & \frac{s^8}{q^2} + \OO{\frac{s^7}{q^2}} - \left( \frac{s^4}{q} + \OO{\frac{s^3}{q}}  \right)^2\\
		& = & \OO{\frac{s^7}{q^2}}
	\end{eqnarray*}
%
\end{proof}

From this deduce 
\begin{proposition}
Let $s \eqdef |\cS|$  and $f$ be such that
$\hf  = \frac{1}{\sqrt{s}} \one_\cS $. For any fixed $\varepsilon >0$ we have as $s$ and $q$ go to infinity
$\prob \left( \left| \sum_{x \in \fq} |f^4(x)| - \frac{s^2}{q^2}\right| > \varepsilon \frac{s^2}{q^2} \right) =o(1)$.
\end{proposition}

\begin{proof}
By Proposition \ref{prop:p2} we know that
\begin{eqnarray*}
	\sum_{x \in \fq} |f^4(x)| &= &\frac{1}{q} \sum_{x \in \fq}  \left|\frac{1}{\sqrt{s}}   \one_\cS \star \frac{1}{\sqrt{s}} \one_\cS(x)\right|^2 \\
	& = & \frac{1}{s^2q} \sum_{x \in \fq} \left|   \one_\cS \star  \one_\cS(x)\right|^2.
\end{eqnarray*}
We then use the Bienaym\'e-Chebyshev inequality with $X \eqdef  \sum_{x \in \fq}  \left|  \one_\cS \star  \one_\cS(x)\right|^2$
$$
\prob\left(|X - \esp(X)| \geq \alpha \sqrt{\var(X)}\right) \leq \frac{1}{\alpha^2}
$$
with $\alpha \eqdef \frac{1}{\sqrt{\var(X)}}\left|\varepsilon \frac{s^4}{q}- \left| \esp(X)-\frac{s^4}{q}\right| \right|$.
We get 
\begin{eqnarray*}
	\prob\left(|X - \frac{s^4}{q}| \geq \varepsilon \frac{s^4}{q}\right) & \leq & 
	\prob\left((|X - \esp(X)| \geq \left|\varepsilon \frac{s^4}{q}- \left| \esp(X)-\frac{s^4}{q}\right|\right|\right) \\
	& = & \prob\left((|X - \esp(X)| \geq \alpha \sqrt{\var(X)}\right)\\
	& \leq & \frac{1}{\alpha^2}.
\end{eqnarray*}
By Lemma \ref{lem:random_S} we know that $\alpha$ tends to infinity with $s$ and $q$, which finishes the proof. 
\end{proof}

\begin{proposition}\label{Proposition:AlgoS}
There exists a quantum polynomial time algorithm that solves $\SAINTERPOL$  with parameters $q,k',S$ as long as $\frac{k'}{q} < 1 - \rho^2$ where $\rho = \frac{s}{q}$, with high probability over a randomly chosen $S \subseteq \F_q$ of size $s$.
\end{proposition}

This is a direct corollary of the previous proposition combined with Theorem~\ref{Theorem:ISISf}.

\section{Benchmarking and comparison of different protocols for $\RSISIS_\infty$ and  $\SAINTERPOL$}
\subsection{Analysis of best classical algorithms for $\SAINTERPOL$}
For simplicity of presentation, we consider the equivalent problem $\SRSC$ (Problem~\ref{Problem:SRSC}). We start from a parity matrix $\Hm' \in \F_q^{k' \times q}$, a set $S \subseteq \F_q$ and a random syndrome $\sv \in \F_q^{k'}$. We define $\rho = \frac{|S|}{q}$. The best algorithm considered in \cite{JSW+24} for solving the OPI problem is basically a variation of the Prange algorithm \cite{P62}. It can also be applied to solve the $\SRSC$ or the $\ISIS_\infty$ problem.  The idea is the following, we fix any vector $\yv_2 \in \F_q^{q-k'}$ such that each coordinate of $\yv_2$ is in $S$, typically, we choose this string at random. Then, we try to find $\yv = \yv_1||\yv_2$ st. $\Hm' \trsp{\yv} = \trsp{\sv}$. This can be done in polynomial time using Gaussian elimination. The resulting $\yv_1 \in \F_q^{k'}$ behaves as a random vector in $\F_q^{k'}$, averaging over $\sv$ and over the choice of $\yv_2$.

\begin{proposition}\cite{P62}
	The above algorithm outputs with overwhelming probability a string $\yv = y_1,\dots,y_n$ st. $|\{i : y_i \in S| \le q\left(1 - \frac{k'}{q}(1-\rho) + o(1)\right)$.
\end{proposition}

This proposition was used in~\cite{JSW+24} to assess the best classical algorithm for this problem\footnote{They also present other classical algorithms in the case we deal with an LDPC code, which is not relevant in our setting.}. This means that unless $\rho = 1 - o(1)$ or $\frac{k'}{q} = o(1)$, there is no regime in which we have a classical polynomial time algorithm for this problem.


\subsection{Comparison of different algorithms: possible rates $R= \frac{k'}{n}$ as a function of $\rho = \frac{|S|}{n}$}

Our main benchmarking criterion is the following: for a fixed constraint size $\rho = \frac{|S|}{q}$, what is the lowest rate $R = \frac{k'}{q}$ for which we can solve $\SRSC$ in polynomial time? Equivalently, we are interested in the smallest degree ratio $R = \frac{k'}{q}$ for which we can find a polynomial of degree $k'$ satisfying the constraints of the $\SAINTERPOL$ problem. We will be interested in regimes $\frac{k'}{q} = \Theta(1)$ and $\rho = \Theta(1)$ and from the discussion above, we don't know any classical polynomial time algorithm that solves $\SRSC$, or equivalently $\SAINTERPOL$ in this regime. 

We now compare the following three algorithms:
\begin{enumerate}
	\item The quantum \emph{Decoded Quantum Interferometry} algorithm from~\cite{JSW+24} for $\SRSC$ with parameter $q,k',S$ that uses the Berlekamp-Welch algorithm. 
	\item  Our quantum algorithm for $\SAINTERPOL$ with parameter $q,k',S$ for a randomly chosen $S$ of a certain size $s$, that uses our reduction and the Koetter-Vardy algorithm.
	\item  Our quantum algorithm for $\RSISIS_\infty$ with parameter $q,k',S = \Iint{-u}{u}$ that uses our reduction and the Koetter-Vardy algorithm.
\end{enumerate}

The second and third algorithm are specified respectively in Proposition~\ref{Proposition:AlgoS} and Theorem~\ref{Theorem:RSISIS}. The Decoding Quantum Interferometry algorithm has the following performance.
\begin{proposition}[\cite{JSW+24}]
	There is a quantum polynomial time algorithm that solves $\SRSC$ with parameters $q,k',S$ as long as $\rho \ge 1 - \frac{k'}{2q}$.
\end{proposition}
This is a direct consequence of (\cite{JSW+24}, Theorem 2.1), which states that there exists a quantum algorithm as long as $\rho \ge 1 - \frac{l}{q}$ where $l < \frac{d^{\perp}}{2}$ where $d^{\perp}$ is the minimal distance of the code $\C$ in which we use our decoding algorithm. The code is a $k$-dimensional Reed-Solomon code with $k = q-k'$ hence its minimal distance $d^{\perp}$ is ${q-k} = {k'}$. This gives $\rho \ge 1 - \frac{k'}{2q}$.

\begin{figure}[!ht]
	\begin{center}
		\includegraphics[width = 10cm]{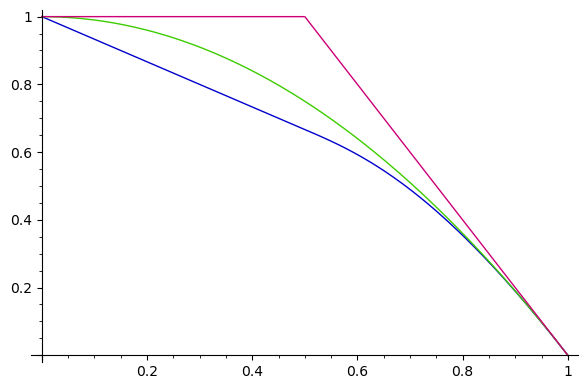} \end{center}
	\caption{Minimum admissible rate $R = \frac{k}{n}$ as a function of $\rho = \frac{|S|}{q}$. The (top) purple line corresponds to the \emph{Decoded Quantum Interferometry} algorithm for $\SAINTERPOL$. The (middle) green line corresponds to our algorithm for $\SAINTERPOL$ with random $S$ and the (bottom) blue line corresponds to  our algorithm with $S = \Iint{-u}{u}$.}
	\label{Figure:MainPlot}
\end{figure}

We put in Figure~\ref{Figure:MainPlot} the smallest value of $R = \frac{k'}{q}$ for which these algorithms run in polynomial time, as a function of $\rho = \frac{|S|}{q}$ . It may seem a little bit odd to compare algorithms for $\SAINTERPOL$ and for $\RSISIS_\infty$ but the best known classical algorithms for these two problems are actually exactly the same, which means that if we are in the quest of a quantum advantage, it is relevant to compare the best quantum algorithms for these problems. Moreover, Decoded Quantum Interferometry also performs equivalently for these two problems.

\textbf{Acknowledgments.} The author thanks Noah Shutty for valuable feedback. We acknowledge funding from the French PEPR integrated projects EPIQ (ANR-22-PETQ-007), PQ-TLS (ANR-22-PETQ-008) and HQI (ANR-22-PNCQ-0002) all part of plan France 2030.


\newpage
\bibliography{paper}
\bibliographystyle{alpha}

\end{document}